\documentclass[12pt,preprint]{aastex}
\usepackage{graphicx}
\begin{document}

\shortauthors{Prieto et~al.}

\title{SN 2008jb: A ``Lost" Core-Collapse Supernova in a Star-Forming Dwarf Galaxy at $\sim 10$~Mpc\altaffilmark{1}}

\author{Jos\'e~L.~Prieto\altaffilmark{2,3,11}, J.~C.~Lee\altaffilmark{2,12}, A.~J.~Drake\altaffilmark{4}, R.~McNaught\altaffilmark{5}, G.~Garradd\altaffilmark{5}, J.~F.~Beacom\altaffilmark{6,7,8}, E.~Beshore\altaffilmark{9}, M.~Catelan\altaffilmark{10}, S.~G.~Djorgovski\altaffilmark{4,14}, G.~Pojmanski\altaffilmark{11}, K.~Z.~Stanek\altaffilmark{7,8}, and D.~M.~Szczygie{\l}\altaffilmark{7}}

\altaffiltext{1}{
 This paper includes data gathered with the 6.5 meter Magellan telescope 
at Las Campanas Observatory, Chile.}

\altaffiltext{2}{Carnegie Observatories, 813 Santa Barbara Street, 
Pasadena, CA 91101, USA} 

\altaffiltext{3}{Department of Astrophysical Sciences, Princeton University, Princeton, NJ 08544, USA}

\altaffiltext{4}{California Institute of Technology, 1200 E. California Blvd., CA 91225, USA}

\altaffiltext{5}{Research School of Astronomy and Astrophysics, Australian National
University, Cotter Road, Weston Creek, ACT 2611, Australia}

\altaffiltext{6}{Department of Physics, The Ohio State University, 191 W. Woodruff Ave., Columbs, OH 43210, USA}

\altaffiltext{7}{Department of Astronomy, The Ohio State University, 140 W. 18th Ave., Columbus, OH 43210, USA}

\altaffiltext{8}{Center for Cosmology and AstroParticle Physics, The Ohio State University, 191 W.\ Woodruff Ave., Columbus, OH 43210, USA}

\altaffiltext{9}{Department of Planetary Sciences, Lunar and Planetary Laboratory, The University of Arizona, 1629 E. University Blvd., Tucson, 
AZ 85721, USA}

\altaffiltext{10}{Departamento de Astronom\'ia y Astrof\'isica, Pontificia Universidad Cat\'olica de Chile, Av. Vicuna Mackena 4860, 782-0436 Macul, 
Santiago, Chile}

\altaffiltext{11}{Warsaw University Astronomical Observatory, Al. Ujazdowskie 4, 00-478 Warsaw, Poland}

\altaffiltext{12}{Hubble, Carnegie-Princeton Fellow}

\altaffiltext{13}{Carnegie Fellow}

\altaffiltext{14}{Distinguished Visiting Professor, King Abdulaziz University, Jeddah 21589, Saudi Arabia}

\begin{abstract}
  We present the discovery and follow-up observations of SN~2008jb, a
  core-collapse supernova in the southern dwarf irregular galaxy
  ESO~302$-$14 \mbox{($M_{B}=-15.3$~mag)} at \mbox{$9.6$~Mpc}. This
  nearby transient was missed by galaxy-targeted surveys and was only
  found in archival optical images obtained by the Catalina Real-time
  Transient Survey and the All-Sky Automated Survey. The well sampled
  archival photometry shows that SN~2008jb was detected shortly after
  explosion and reached a bright optical maximum, $V_{\rm max} \simeq
  13.6$~mag ($M_{V,\rm max}\simeq -16.5$). The shape of the light
  curve shows a plateau of $\sim 100$~days, followed by a drop of
  $\sim 1.4$~mag in $V$-band to a slow decline with the approximate
  $^{56}$Co decay slope.  The late-time light curve is consistent with
  $0.04\pm 0.01$~M$_{\odot}$ of $^{56}$Ni synthesized in the
  explosion. A spectrum of the supernova obtained 2~years after
  explosion shows a broad, boxy H$\alpha$ emission line, which is
  unusual for normal type~IIP supernovae at late times. We detect the
  supernova in archival Spitzer and WISE images obtained $8-14$~months
  after explosion, which show clear signs of warm ($600-700$~K) dust
  emission. The dwarf irregular host galaxy, ESO~302$-$14, has a low
  gas-phase oxygen abundance, $\rm 12 + log(O/H) = 8.2$ ($\sim
  1/5$~Z$_{\odot}$), similar to those of the SMC and the hosts of long
  gamma-ray bursts and luminous core-collapse supernovae. This
  metallicity is one of the lowest among local ($\lesssim 10$~Mpc)
  supernova hosts. We study the host environment using GALEX far-UV,
  $R$-band, and H$\alpha$ images and find that the supernova occurred
  in a large star-formation complex. The morphology of the H$\alpha$
  emission appears as a large shell ($R \simeq 350$~pc) surrounding
  the FUV and optical emission. Using the H$\alpha$-to-FUV ratio, and
  FUV and $R$-band luminosities, we estimate an age of $\sim 9$~Myr
  and a total mass of $\sim 2\times 10^{5}$~M$_{\odot}$ for the
  star-formation complex, assuming a single-age starburst. These
  properties are consistent with the expanding H$\alpha$ supershells
  observed in many well-studied nearby dwarf galaxies, which are
  tell-tale signs of feedback from the cumulative effect of massive
  star winds and supernovae. The age estimated for the star-forming
  region where SN~2008jb exploded suggests a relatively high-mass
  progenitor star with initial mass $\rm M \sim 20$~M$_{\odot}$, and
  warrants further study. We discuss the implications of these
  findings in the study of core-collapse supernova progenitors.
\end{abstract}


\section{Introduction}

Core-collapse supernovae are energetic explosions that mark the death
of stars more massive than $\approx 8-10$~M$_{\odot}$. They are
extremely important in several areas of astrophysics, including the
nucleosynthesis of chemical elements, energy feedback that affects the
evolution of galaxies, the formation of compact object remnants, the
production of high-energy particles, as tracers of recent
star-formation in galaxies, and as test cases of massive stellar
evolution.

Supernovae discovered in nearby galaxies have been particularly
important for testing our physical understanding of the explosions and
to establish direct links between the events and their progenitors,
which puts tight constraints on stellar evolution theory. The two
best-studied nearby core-collapse supernovae, the peculiar type~II
SN~1987A in the LMC and the type~IIb SN~1993J in M81, had progenitors
that were detected in pre-explosion images. Both were very interesting
cases that challenged theoretical expectations based on stellar
evolution models that predicted red supergiants as their progenitor
stars: SN~1987A had a $\simeq 20$~M$_{\odot}$ blue supergiant
progenitor star (e.g., Arnett et al. 1989) and SN~1993J had a $\simeq
15$~M$_{\odot}$ yellow supergiant progenitor (e.g., Aldering et
al. 1994), which turned out to be a massive binary system (e.g., Maund
et al. 2004). After more than a decade of work, Smartt et al. (2009)
presented a thorough study of nearby ($\lesssim 30$~Mpc)
type~II-Plateau supernovae with deep pre-explosion imaging (see also
Li et al. 2005), which allowed them to constrain the progenitors of
type~IIP supernovae to be red supergiants with initial main-sequence
masses in the range $8-17$~M$_{\odot}$. This result is in conflict
with well-studied red supergiant samples in the Galaxy and the
Magellanic Clouds (e.g., Levesque et al. 2006), which also contain
massive supergiants with $\rm M\simeq 20-30$~M$_{\odot}$, and a
Salpeter Initial Mass Function (IMF) for high-mass stars.

This ``red supergiant problem" can be alleviated if high-mass red or
yellow supergiants produce other spectroscopic or photometric type~II
supernova sub-types (e.g., Smith et al. 2009; Elias-Rosa et al. 2009,
2010), or if high-mass supergiants experience strong winds due to
pulsations (e.g., Yoon \& Cantiello 2010). Very recently, Walmswell \&
Eldridge (2011) have proposed circumstellar dust as a possible solution.
It can also be explained by failed supernovae, an hypothesis that would
have several deep implications (e.g., Kochanek et al. 2008). In
addition, there are known selection biases that have not been discussed
in depth or accounted for. Of the 18 nearby type~II supernovae with
fairly secure progenitor (either single, binaries, or compact clusters)
detections\footnote{These are: SN~1987A, 1993J; SN~1961V from Kochanek
et al. (2011) and Smith et al. (2011c); SN~1999ev, 2003gd, 2004A,
2004am, 2004dj, 2004et, 2005cs, and 2008bk from the sample of Smartt et
al. (2009); SN~2005gl, from Gal-Yam et al. (2007) and Gal-Yam \& Leonard
(2009); SN~2008ax from Crockett et al. (2009); SN~2008cn and 2009kr from
Elias-Rosa et al.  (2009, 2010); SN~2009md from Fraser et al. (2011);
SN~2010jl from Smith et al. (2011b); and SN~2011hd from Maund et al.
(2011), Szczygiel et al. (2011) and Van Dyk et al. (2011). We do not
include here low-luminosity transients with type~IIn-like spectral
features and progenitor detections in the mid infrared, like SN~2008S
and NGC~300-OT (e.g., Prieto et al. 2008b).}, two of the 18 events were
initially discovered by the galaxy-targeted Lick Observatory Supernova
Search (LOSS; Filippenko et al. 2001), and 14 of the 18 were initially
discovered by dedicated amateur astronomers that mostly concentrate
their search efforts in big galaxies. Only 2 of these supernovae were
found in dwarf galaxies (SN~1987A and 2008ax) and all the type~IIPs (10
of the 18) were found in the disks of spiral galaxies, mostly in
grand-design spirals.

The existing selection bias against finding nearby core-collapse
supernovae in dwarf galaxies could be important because $20-40\%$ of
the local star formation rate density is in galaxies with absolute
magnitude $M_B > -18$~mag (e.g., James et al. 2008; Young et al. 2008;
Williams et al. 2011).  Since metallicity is a key parameter in
massive stellar evolution and stellar death (e.g., Prieto et
al. 2008a; Modjaz 2011, and references therein), changes in the
fractions of different type~II spectroscopic subtypes with metallicity
(e.g., Arcavi et al. 2010) could affect our census of progenitors in
local samples. Also, possible variations of the stellar IMF with
environment (e.g., Meurer et al. 2009; but see, e.g., Myers et
al. 2011 and references therein) can introduce complications to the
progenitor analyses. Recently, Horiuchi et al. (2011) discussed in
detail related effects of incompleteness in the existing supernova
rate measurements and how they affect their use as star formation rate
indicators.
 
In this paper we present the discovery, follow-up observations, and
analysis of the nearby type~II SN~2008jb ($\alpha = 03^{\rm h}51^{\rm
  m}44\fs 66$, $\delta = -38^{\circ} 27\arcmin 00\farcs1$) in a
southern dwarf-irregular galaxy at $\sim 10$~Mpc (Prieto et
al. 2011). This bright $13.6$~mag supernova was missed by targeted
southern supernova surveys, like CHASE (Pignata et al. 2009) and
amateur searches, mainly because the host galaxy ESO~302$-$14 is a
fairly low-luminosity dwarf and was not included in the catalogs of
galaxies that are surveyed for supernovae. We are able to recover the
optical light curves of this supernova from two surveys that do not
target individual galaxies, but rather scan large areas of the sky:
the Catalina Real-time Transient Survey\footnote{{\tt
    http://crts.caltech.edu}} (CRTS; Drake et al. 2009; Djorgovski et
al. 2011) and the All-Sky Automated Survey (ASAS; Pojmanski 1998). In
Section~\S\ref{sec1} we discuss the observations, including the
optical discovery and follow-up, archival mid-infrared imaging,
optical spectroscopy, and other archival data. In Section~\S\ref{sec2}
we present detailed analysis and discussion of the optical and mid-IR
observations. In Section~\S\ref{sec3} we discuss the results and
conclusions. We adopt a distance of 9.6~Mpc to ESO~302$-$14 (Lee et
al. 2009), estimated from the Virgo-inflow corrected recession
velocity and $H_0=75$~km~s$^{-1}$~Mpc$^{-1}$.

\section{Observations}
\label{sec1}

\subsection{Optical Imaging}

SN~2008jb was discovered in CRTS archival data from the Siding Spring
Survey (SSS) 0.5m Schmidt telescope through the SNhunt
project\footnote{{\tt
    http://nesssi.cacr.caltech.edu/catalina/current.html}}. SNhunt is
an open survey for transients in nearby galaxies that uses image
subtraction in images obtained by the CRTS. A new transient was
recovered in 2010 Oct. 26 (ID SNhunt12) after running the difference
imaging pipeline in archival data from the SSS. SN~2008jb is first
detected in unfiltered SSS images obtained on 2008 Nov. 23, at $V
\simeq 13.6$~mag. Figure~1 shows the first detection of SN~2008jb in
the SSS images. It was also detected in 10 images obtained between
2008 Dec. 3 and 2010 Jan. 11. We used the image subtraction software
described in Freedman et al. (2009) in order to obtain clean images of
the supernova (see Fig.~1). The image used as a template for the
subtractions was a combination of images obtained in 2007. We
performed PSF photometry using DAOPHOT (Stetson 1992) on the
difference images. The final calibration of the supernova photometry
was done relative to ASAS $V$-band data of stars in the field (see
discussion below). Table~1 gives the $V$-band calibrated photometry
from the SSS data. We also include 3$\sigma$ upper limits on the
magnitudes of the supernova obtained from the images just before the
first detection and after the last detection.
 
The supernova was also detected in archival images collected using the
7~cm ASAS South telescopes in Las Campanas Observatory, Chile. The
first detection in ASAS South data is from 2008 Nov. 19 (4 days
earlier than CRTS) at $13.9$~mag in the $I$-band. The ASAS images were
processed with the reduction pipeline described in
\citet{pojmanski98,pojmanski02}. The $V$ and $I$ magnitudes are tied
to the Johnson $V$ and Cousins $I$ scale using Tycho \citep{hog00} and
Landolt \citep{landolt83} standard stars. We used aperture photometry
(2~pixel radius) from the normal reduction pipeline to obtain the
magnitudes of the supernova in ASAS images and estimated $3\sigma$
upper limits before and after the first and last detection epochs,
respectively. Unlike the case of the SSS images where there is
contamination from the nearby star-forming region at faint magnitudes,
the much shallower ASAS data ($V\lesssim 15$~mag) does not require
image subtraction. The ASAS $V$ and $I$ magnitudes of SN~2008jb are
presented in Table~2. Figure~2 shows the light curve of the supernova
from ASAS and CRTS data.

We obtained late-time observations of SN~2008jb on 2010 Nov. 7 using
the Inamori-Magellan Areal Camera \& Spectrograph (IMACS; Dressler et
al. 2006) on the 6.5~m Magellan I (Baade) telescope at Las Campanas
Observatory. We obtained $3\times 200$~sec images of the field of
SN~2008jb with the f/2 camera ($0\farcs2$ per pixel) using the
$R$-band filter under good weather and seeing conditions ($0\farcs7 -
0\farcs8$). Since we lack the deep template image necessary to obtain
the supernova flux at late times, these data are only used to study
the host galaxy environment. We plan to obtain more late-time
observations and the analysis of these data will be presented in a
future publication.

\subsection{Optical Spectroscopy}

We gathered spectroscopic observations of SN~2008jb on 2011 Jan 6
using the IMACS on the 6.5~m Baade telescope at Las Campanas
Observatory. The observations consisted of $3\times 1800$~sec spectra
obtained using the 300~l/mm grism (range $4000-9000$~\AA) and a
$0\farcs9$ slit aligned at the parallactic angle, obtained under clear
weather and good seeing ($0\farcs7$). This setup gives a FWHM
resolution of $4$~\AA\ at 6500~\AA. The data were reduced using
standard techniques in IRAF, which included basic data reduction
(overscan and bias subtraction, flat-fielding), 1D spectrum
extraction, wavelength calibration using a HeNeAr arclamp obtained
after the science observations, and flux calibration using a flux
standard observed on the same night.  The spectrum goes through two CCD
detectors separated by a gap of $\sim 74$~\AA. We used a simple linear
interpolation at the edges of the CCDs. The final spectrum of
SN~2008jb is shown in Figure~3.

We also obtained the spectrum of an \ion{H}{2} region in the host galaxy
at $\sim 150$~pc ($\sim 3\arcsec$) south of the supernova position on
2011 Jan. 2 with the Wide-Field Reimaging CCD Camera (WFCCD) on the
2.5-m du Pont telescope at Las Campanas Observatory. The spectrum
consisted of $3\times 1800$~sec exposures with the 400~l/mm grism, which
gives a FWHM resolution of 8~\AA\ and continuous coverage in the
$3800-9200$~\AA\ range. The data were reduced using standard tasks in
IRAF similar to the reduction procedure of the IMACS supernova spectrum.
The final spectrum of the \ion{H}{2} region is shown in Figure~4. We
measured the fluxes of the most prominent emission lines in the spectrum
using Gaussian profiles and a custom Perl Data Language (PDL) fitting
routine, including the Balmer recombination lines (H$\beta$
$\lambda$~4862 and H$\alpha$ $\lambda$~6563) and forbidden emission
lines ([\ion{O}{3}] $\lambda\lambda$~4959, 5007, [\ion{S}{2}]
$\lambda\lambda$~6713, 6731, and \mbox{[\ion{N}{2}] $\lambda$~6583}).
The fluxes of the optical emission lines are presented in Table~3.

\subsection{Mid-Infrared Imaging}

The host galaxy of SN~2008jb, ESO~302$-$14, was observed with
Spitzer/IRAC (Fazio et al. 2004) using the 3.6 and 4.5~$\mu$m channels
in warm Spitzer observations obtained as part of the Spitzer Survey of
Stellar Structure in Galaxies (S4G; Sheth et al. 2010). They obtained
two epochs of imaging during Cycle 6 (PID 61060) separated by $\sim
1$~month, in 2009 Sep. 2 and Oct. 10. We retrieved the post-BCD images
of ESO~302$-$14 from the Spitzer Heritage Archive, which are fully
reduced and flux-calibrated. The supernova is clearly detected as a
bright, variable mid-IR source in these images, at a position
consistent with the optical coordinates. The mid-IR detections are
shown in Figure~5. We measured the 3.6 and 4.5~$\mu$m fluxes of the
source using aperture photometry with a $2\farcs4$ aperture radius
($2\farcs4 - 7\farcs2$ annulus for sky determination) and applied the
aperture corrections and flux conversion factors listed in the IRAC
Instrument Handbook. The IRAC fluxes are presented in Table~4.

The Wide-field Infrared Survey Explorer (WISE; Wright et al. 2010)
preliminary data release includes observations of the field of
SN~2008jb obtained at 3.4, 4.6, 12, and 22~$\mu$m between 2010 Jan. 23
and Jan. 30. We searched for the WISE data in the NASA/IPAC Infrared
Science Archive (IRSA). The supernova is clearly detected at 3.4 and
4.6~$\mu$m (see Figure~5), marginally detected at 12~$\mu$m, and
undetected at 22~$\mu$m. We retrieved the photometry of source
J035144.63-382700.6 from the WISE catalog and list the fluxes in
Table~4. We did not attempt to measure photometry in individual WISE
images, but rely on the catalog fluxes measured from the combined
data. The effective date of the observations is 2010 Jan. 27.

\subsection{Other Archival Data}

We searched for other existing archival data of ESO~302$-$14 obtained
before the discovery of SN~2008jb to help characterize the progenitor
and galaxy environment. The Survey for Ionization in Neutral Gas
Galaxies (SINGG; Meurer et al. 2006) obtained $R$-band and H$\alpha$
narrow-band images of the host galaxy using the CTIO 1.5-m telescope
on 2000 Oct. 28 (seeing was $1\arcsec$ in $R$). We retrieved
calibrated images from SINGG (including H$\alpha$-subtracted images)
through the NASA Extragalactic Database (NED).

The Galaxy Evolution Explorer (GALEX; Martin et al. 2005) ultraviolet
space telescope observed the field of ESO~302$-$14 on 2004 Nov. 18.
We retrieved calibrated NUV and FUV images (GR6 data release) from 
the GALEX online archive.

We also searched for pre-explosion imaging data in the {\it HST},
Gemini, and ESO archives. Unfortunately, there are no deep
pre-explosion images obtained with these facilities.

\section{Analysis}
\label{sec2}

In the following sections we present analysis and discussion of the
results from the light curves, spectra, mid-IR emission, and host
galaxy environment. We present a summary of derived physical
properties obtained from the optical light curves and host galaxy in
Table~5 and Table~6, respectively.

\subsection{Light Curve}

In order to make light curve comparisons and derive intrinsic physical
properties, we first need to estimate the total reddening along the
line of sight to the supernova. For the Galactic reddening, we use a
CCM reddening law with $R_{V}=3.1$ (Cardelli, Clayton \& Mathis 1989)
and $E(B-V)_{\rm MW}=0.009$~mag from the reddening maps of Schlegel et
al. (1998). We estimate the intrinsic reddening in the host galaxy
using the Balmer decrement measured from the spectrum of the
\ion{H}{2} region in ESO~302$-$14 (see Table~3). We assume an
intrinsic case B recombination Balmer flux ratio $\rm H\alpha/H\beta =
2.86$, which is appropriate for an \ion{H}{2} region at a typical
electron temperature and density (Storey \& Hummer 1995). Then we
assume an SMC reddening law from Gordon et al. (2003) for the host
galaxy, which is appropriate given its star-formation rate, absolute
magnitude, and metallicity (see Section~3.4). The resulting intrinsic
color excess is $E(B-V)_{\rm host}=0.06\pm 0.02$~mag and extinction
$A_{V}=0.16\pm 0.06$~mag, which we will use in all subsequent
analysis. The values obtained using an LMC reddening law, $E(B-V)_{\rm
  host}=0.05$~mag and $A_{V}=0.17$~mag, and Galactic CCM reddening
law, $E(B-V)_{\rm host}=0.06$~mag and $A_{V}=0.19$~mag, are consistent
with our adopted reddening. We note that for \ion{H}{2} regions its
usually assumed $E(B-V)_\star = 0.44\,E(B-V)_{\rm gas}$ (Calzetti
2000), but we adopt conservatively $E(B-V)_{\rm host} \simeq
E(B-V)_{\rm gas}$ because of the young age of the region (see
Section~3.4).

The $V$- and $I$-band light curves of SN~2008jb resemble those of
type~II-Plateau supernovae, the most common kind of core-collapse
events in nearby galaxies (e.g., Li et al. 2011). In Figure~6 we
compare the absolute $V$-band light curves of different core-collapse
supernovae with SN~2008jb, including: the low-luminosity type~IIP
2005cs (Pastorello et al. 2009a), the luminous type~IIP 2004et
(Maguire et al. 2010), and the low-luminosity type~IIL 1999ga
(Pastorello et al. 2009b). The absolute magnitudes of SN~2008jb lie in
between SN~2005cs and SN~2004et, and appear consistent with SN~1999ga
and normal type~IIP supernovae (e.g., Li et al. 2011). In the
comparison we have adopted the host distances, total extinctions, and
explosion times presented in the published studies. We estimate an
approximate explosion time for SN~2008jb as the midpoint between the
last pre-discovery $V$-band non-detection from CRTS and the first
$I$-band detection from ASAS, which gives $\rm HJD_{exp}\simeq
2454782.0$ (2008 Nov. 11).

The ASAS data samples well the initial plateau, which lasts $\sim
95$~days in the $I$-band. This plateau duration is in the observed
range of plateaus in type~IIP supernovae, typically between $\sim
80-120$~days (e.g., Bersten \& Hamuy 2009). The $I$-band light curve
shows a slow linear decay of $0.007$~mag/day and the $V$-band light
curve declines faster at $0.013$~mag/day in the initial phase, which
is slightly faster than well-studied type~IIP supernovae where the
initial decay slope in the plateau is slower at redder bands (e.g.,
Poznanski et al. 2009; D'Andrea et al. 2010). It is interesting to
note the similarities of the initial light curve decline of SN~2008jb
with the light curve of the low-luminosity type~IIL SN~1999ga
(Pastorello et al. 2009b), as shown in Figure~6. The $V$-band light
curves of type IIL supernovae typically have faster initial decline
slopes than SN~2008jb and SN~1999ga (e.g., Barbon et al. 1979; Fig. 3
of Pastorello et al. 2009b).

The absolute magnitudes close to the explosion date are $M_{V,\rm
  max}\simeq -16.52$~mag and $M_{I,\rm max}\simeq -16.65$~mag. At mid
plateau ($\sim 50$~days), the absolute magnitudes are $M_{V,\rm
  mid}\simeq -15.87$~mag and $M_{I,\rm mid}\simeq -16.37$~mag. The
evolution of the $V-I$ color, from $(V-I)_{0}\simeq 0.13$~mag close to
explosion to $(V-I)_{0}\simeq 0.50$~mag at mid plateau, is also fairly
consistent with the evolution of well-studied type~IIP supernovae,
which traces small changes in the color temperature of the photosphere
(e.g., Hamuy et al. 2001). It is worth noting that the unreddened
color at mid-plateau of SN~2008jb is very close to the $(V-I)_0 =
0.53$~mag ridgeline that Nugent et al. (2006) derived from a large
sample of type~IIP supernovae, supporting our reddening estimation.

After the $\sim 100$~day plateau, the light curve drops $\sim 1.4$~mag
in 32~days (at $\sim 0.044$~mag/day) to a late-time linear decay of
$0.013$~mag/day. We do not have $I$-band observations that show the
transition phase, but the ASAS non-detections at these epochs are
consistent with the $V$-band light curve shape. The $V$-band drop to a
late-time linear decay is seen in all type~IIP supernovae, but it is
typically stronger ($\gtrsim 2$~mag in the $V$-band; e.g., Maguire et
al. 2010). In Figure~6 we show that the late-time linear decay of
SN~2008jb is similar to other type~II supernovae and reasonably
consistent with the $^{56}$Co to $^{56}$Fe radioactive decay slope.

Hamuy (2003) studied the physical properties of a sample of
well-observed type~II supernovae and derived the $^{56}$Ni masses
produced in the explosions from their late-time light curves, assuming
full trapping of $\gamma$-ray photons produced by the radioactive
decay of $^{56}$Co.  Using Eq.~1 in his study evaluated at day 115
after explosion (the first light curve point at the radioactive tail)
we obtain a $^{56}$Ni mass of $0.04 \pm 0.01$~M$_{\odot}$. This result
assumes a bolometric correction of $0.26 \pm 0.06$~mag in the
$V$-band, which is calculated from the well-studied SN~1987A and
SN~1999em at nebular phases (Hamuy 2003). The estimated uncertainty
does not include a systematic error in the distance to the host.

\subsection{Spectrum}

The late-time spectrum of SN~2008jb, obtained $\sim 2$~years after the
explosion, shows a prominent H$\alpha$ emission feature (see
Figure~3), supporting the classification of this transient as a
type~II. The H$\alpha$ feature is broad ($\rm EW \sim 480$~\AA),
flat-topped, boxy, and blueshifted (centered at $\sim
-1300$~km~s$^{-1}$), with the blue edge of the line at
$-8700$~km~s$^{-1}$ and red edge at $7400$~km~s$^{-1}$ after correcting
for the host galaxy's recession velocity. The spectrum does not show any
other strong emission features characteristic of normal type~II
supernovae at late times (e.g., [\ion{O}{1}] and [\ion{Ca}{2}]), except
for a tentative low S/N feature centered at $6180$~\AA\ with $\rm FWHM
\sim 3600$~km~s$^{-1}$. If the spectrum is heavily smoothed, there is a
``bump'' that shows up more clearly at $\gtrsim 7000$~\AA, which could
correspond to low-level emission associated with the [\ion{Ca}{2}]
doublet at $\lambda\lambda$~7291, 7323. There is an increase in flux to
the blue of $\sim 5500$~\AA, which could in part be explained by
contamination from nearby sources, although the S/N in the blue
part of the spectrum decreases also due to flatfielding errors.

The broad and boxy H$\alpha$ emission feature detected in SN~2008jb is
not typically observed in late-time spectra of type~II supernovae. We
used spectra of normal type~IIPs in the SUSPECT database to measure
the FWHM at $\gtrsim 300$~days after explosion, in the nebular phase,
and measure H$\alpha$ line widths of $\sim 2300-2900$~km~s$^{-1}$,
significantly lower velocities than SN~2008jb. The peculiar SN~2007od
(Andrews et al. 2010; Inserra et al. 2011) is the only example we
could find in the literature of a type~IIP with a comparably broad and
blueshifted H$\alpha$ component at late-times. In this case, however,
the line also showed multiple narrower peaks on top of the broad
profile that indicated clear signs of interaction between the ejecta
and the circumstellar medium (CSM) from the projenitor wind, which are
not seen in SN~2008jb.

Other less common type~II supernovae have shown very broad emission
lines at late times. The type~IIL SN~1979C in M100 (e.g., Branch et
al. 1981) showed a broad $\sim 15000$~km~s$^{-1}$ H$\alpha$ profile
one year after discovery (Cappellaro et al. 1995). Also the type~IIL
SN~1980K showed an H$\alpha$ profile with $\rm FWHM \sim
10000$~km~s$^{-1}$ eight years after the explosion (Fesen \& Becker
1990), although at earlier epochs comparable to our observations of
SN~2008jb the line widths were significantly narrower (e.g., Capellaro
et al. 1995). The well-studied type~IIb SN~1993J showed a strong
H$\alpha$ emission feature at $>1$~year after explosion, with velocity
and profile shape fairly similar to SN~2008jb (Filippenko et al. 1994;
Matheson et al. 2000). The boxy feature observed in SN~1993J was
interpreted as indication of circumstellar interaction in a spherical
shell (Matheson et al. 2000), which supported the evidence from the
radio and X-ray observations (e.g., Fransson et al. 1996 and
references therein). Patat et al. (1995) had also shown that the
luminosity of the H$\alpha$ line $0.5-1$~year after the explosion was
in excess of the expectations from radioactive $^{56}$Co decay, and
the most likely source of extra energy was the ejecta-CSM interaction.

Chugai (1991) presented models of the evolution of the H$\alpha$ line
luminosity as a function of time for type~II supernovas assuming
radioactive decay as the energy source. We used these models to test
if the measured luminosity of the H$\alpha$ emission line in SN~2008jb
at $\sim 800$~days after explosion is consistent with heating from
radioactive decay or extra energy is needed to explain it. We use the
parametrization of the models presented in Pastorello et al. (2009b;
their Fig.~9), scaled to a nickel mass of $0.04$~M$_{\odot}$, and
extrapolated the curves linearly at late-times (in log~L versus time).
Figure~7 shows the models with different assumptions for the total
ejected mass from the progenitor ($\rm M=5, 14, 20$~M$_{\odot}$) and
the measured H$\alpha$ luminosity of SN~2008jb after correcting for
Galactic and internal extinction. There is a clear excess in H$\alpha$
emission luminosity with respect to the models at $\sim 800$~days by a
factor of $\sim 1.5$, suggesting the existence of an extra energy
source (e.g., from ejecta interaction with the progenitor wind) or a
more massive progenitor star ($\gtrsim
20$~M$_{\odot}$). Unfortunately, we do not have early spectra to trace
the evolution in H$\alpha$ emission luminosity with time to
differentiate between these scenarios. Also, the gap between CCD chips
is right at the wavelength of the H$\alpha$ line, so we are unable to
detect a narrow component that would be a clear indication of
circumstellar interaction. 

\subsection{Mid-IR Emission}

We clearly detect mid-IR emission from SN~2008jb in three Spitzer and
WISE epochs obtained between 286 and 434 days after the first optical
detection (see Figure~5). The red $\rm [3.6] - [4.5]$ Spitzer color of
$1.2$~mag indicates a rising spectral energy distribution at mid-IR
wavelengths, a clear sign of warm dust emission. In order to
characterize the evolution of the optical and mid-IR emission, we fit
the SED of the supernova using two black-bodies, a hot component to
fit the optical $V$-band data and a warm component to fit the mid-IR
data. Since we only have single-band optical data at late-times, we
assume an effective temperature of $\rm T_{eff} = 6000$~K for the hot
component, which is a typical temperature measured in normal type~II
supernovae in the nebular phase.

The results of the two component black-body fits are presented in
Table~7, and Figure~8 shows the model fits to the SED as a function of
time. The total integrated luminosity decreases by a factor of 6.8 in
$148$~days, which is equivalent to 0.014~mag/day. The luminosity
contributed by the warm black-body component, which is better
constrained from the observed SED, decreases by a factor of 5.5 in
148~days, or $0.013$~mag/day. The consistency with the late-time
$V$-band decline and the $^{56}$Co decay slope is interesting and
argues for radioactive decay as the dominant energy source of dust
heating. The temperature of the warm black-body component is
$600-700$~K in the three epochs where we have mid-IR imaging. The WISE
data from day 434 clearly shows that the SED peaks between 4.6 and
12~$\mu$m, supporting our conclusion. We also find that the radius of
the warm black-body component decreases by a factor of $\sim 3$
between the first and last epochs with mid-IR data, between 765~AU
(day 286) and 237~AU (day 434).  The total mid-IR luminosity can be
used to estimate the amount of dust needed to explain it (e.g., Dwek
et al. 1983; Prieto et al.  2009); we find a dust mass of $\sim
10^{-5} - 10^{-4}$~M$_{\odot}$ assuming a black-body spectrum and
carbon dust composition.

Several nearby type~II supernovae have shown excess mid-IR emission at
late times, which is interpreted as the presence of warm dust either
newly formed or pre-existing in the progenitor CSM (or a
combination). Some examples of nearby type~IIP events with clear signs
of warm dust emission include SN~2003gd (Sugerman et al. 2006; Meikle
et al. 2007), SN~2004et (Kotak et al. 2009), SN~2004dj (Szalai et
al. 2011; Meikle et al. 2011), SN~2007it (Andrews et al. 2011), and
SN~2007od (Andrews et al. 2010).  The classic, luminous type~IIL
SN~1979C and SN~1980K showed infrared excesses at late times (Dwek
1983; Dwek et al. 1983).  Also, a large fraction of type~IIn
supernovae show mid-IR emission at late times, which has been
associated with pre-existing CSM dust (e.g., Fox et al. 2011, and
references therein). In the low-luminosity end, SN~2008S-like events
have dusty massive star progenitors and also show signs of dust during
the transients (e.g., Prieto et al. 2008b, 2009; Thompson et al. 2009;
Kochanek 2011).

The typical dust masses needed to explain the mid-IR emission in
normal type~II supernovae are $\lesssim 10^{-3}$~M$_{\odot}$ (but see
Matsuura et al. 2011 for the discovery of a large reservoir of cold
dust in SN~1987A) and fairly consistent with the range derived here
for SN~2008jb. The observations for several of the well-studied
type~IIs have been usually explained by newly formed dust in the
ejecta. It seems unlikely that this can explain SN~2008jb. The main
reason for this is that the shock velocities inferred from the
black-body fits to the mid-IR detections are in the range of
$1000-4500$~km~s$^{-1}$ (depending on the epoch), which is quite low
compared to the observed velocity of the H$\alpha$ emission line. On
the other hand, a decreasing black-body radius as a function of time
has been seen in SN~2008S and NGC~300-OT (also the type~IIP SN~2007it
and SN~2007od), and has been explained by Kochanek (2011) using a
model in which dust reforms in the progenitor wind. This model,
however, requires very high densities and low velocities, which are
not observed in SN~2008jb. 

\subsection{Host Galaxy Environment}

The host of SN~2008jb, ESO~302$-$14, is a star-forming irregular
galaxy similar to the Magellanic Clouds. In particular, it has a total
$B$-band absolute magnitude \mbox{$M_B = -15.3$~mag}, $\sim 1$~mag
fainter than the SMC, with a low UV derived (total) current
star-formation rate (SFR) of 0.03~M$_{\odot}$~yr$^{-1}$ (Lee et
al. 2009). The total stellar mass of the galaxy is $4\times
10^7$~M$_{\odot}$, estimated from $M_B$, $M_R$, and the mass-to-light
ratios as a function of $B-R$ presented in Bell \& de~Jong (2001),
assuming a Salpeter IMF. The total neutral hydrogen (\ion{H}{1}) mass
is $3\times 10^{8}$~M$_{\odot}$ (Meurer et al. 2006, scaled by our
adopted distance). These masses imply high SFRs per unit stellar mass,
$\rm log(SFR/M_{\star}) \simeq -9.1$ and $\rm log(SFR/M_{H\,\sc{I}})
\simeq -9.9$, which are typical for local star-forming dwarf galaxies
(e.g., Lee et al. 2011).

We use the spectrum we obtained of the \ion{H}{2} region at $\sim
150$~pc from the position of SN~2008jb to measure the local oxygen
abundance using the strong forbidden oxygen lines and hydrogen
recombination lines. Using the emission line fluxes reported in
Table~3 and correcting for host galaxy extinction, we estimated the
[\ion{O}{3}]~$\lambda$5007/H$\beta$ and [\ion{N}{2}]/H$\alpha$ line
ratios. Then we used the $N2$ and $O3N2$ calibrations from Pettini \&
Pagel (2004; PP04) to estimate oxygen abundances of $\rm 12+log(O/H) =
8.13 \pm 0.03$ ($N2$) and $8.21 \pm 0.03$ ($O3N2$). These
uncertainties are statistical and do not include the $0.1-0.2$~dex
error in the oxygen abundance calibration (PP04). For comparison, the
average oxygen abundance of \ion{H}{2} regions in the SMC from Russell
\& Dopita (1990) are $\rm 12 + log(O/H) = 8.08$ ($N2$) and $8.11$
($O3N2$).

The environment of SN~2008jb in ESO~302$-$14 has a low oxygen
abundance compared to the environments of nearby type~IIP supernovae
($\lesssim 30$~Mpc) used to constrain progenitor properties from deep
the pre-explosion observations (Smartt et al. 2009), although there
are also normal core-collapse supernovae discovered in relatively
nearby galaxies that have fairly low-metallicity environments (e.g.,
Prieto et al. 2008b; Anderson et al. 2010). The metallicity here is
similar to the environments of long-duration GRB hosts (e.g., Stanek
et al. 2006; Levesque et al. 2010) and the hosts of the most luminous
core-collapse supernovae that are being discovered in galaxy-blind
surveys (e.g., Kozlowski et al. 2010; Stoll et al. 2011), which are
bound to contain a population of very massive stars.

By closely examing the local host galaxy properties we can constrain
the progenitor properties (see, e.g., Anderson \& James 2008,
2009). We have relatively deep optical images obtained at late times
with Magellan/IMACS, archival UV data from GALEX and H$\alpha$ from
SINGG.  Figure~9 shows a mosaic with the Magellan $R$-band and GALEX
FUV image, scaled to fit the whole galaxy (top panels) and the region
where the supernova exploded (lower panels). We see that SN~2008jb
exploded in a large star-formation complex, the brightest and highest
surface brightness star-forming region within the galaxy in the
optical and FUV. It is composed of at least two resolved clusters or
stellar associations that are well separated in the Magellan $R$-band
image. Interestingly, the H$\alpha$ emission (red contours in the
lower panel of Fig.~9) is offset and outside the brightest optical and
FUV emission, forming an apparent ring with a projected diameter of
$\simeq 700$~pc.

We can estimate an approximate age for the star-forming complex in
ESO~302$-$14 using the ratio of H$\alpha$ to FUV luminosities, which
is a sensitive age indicator in an instantaneous burst of star
formation (e.g., Stewart et al. 2000; S{\'a}nchez-Gil et al. 2011). We
use the Starburst99 (Leitherer et al.  1999) models to generate a grid
of single-age clusters with total masses between $5\times
10^4$~M$_{\odot}$ and $4\times 10^6$~M$_{\odot}$, in steps of
$10^5$~M$_{\odot}$. We choose a Salpeter IMF with stellar masses
between 0.1 and 100~M$_{\odot}$ and Geneva stellar evolution models
with $Z=0.004$, consistent with the measured oxygen abundance of the
\ion{H}{2} region. Figure~10 shows the expected ratio of
H$\alpha$-to-FUV luminosities as a function of the FUV luminosities
for these models. We label the ages of the clusters between 1~Myr and
50~Myr. The total measured H$\alpha$-to-FUV ratio $\rm
log(L_{H\alpha}/L_{FUV}) = 12.41\pm 0.16$ and FUV luminosity $\rm
L_{FUV} = (5.7 \pm 1.8) \times 10^{25}$~erg~s$^{-1}$~Hz$^{-1}$ of the
star-forming complex, corrected by Galactic and host galaxy
extinction, are shown in the figure. We obtain an age of $8.8\pm
0.8$~Myr and total mass of $\rm (2.4\pm 1.0) \times
10^{5}$~M$_{\odot}$. We also used the integrated $R$-band flux
(extinction corrected) of the star-forming region to independently
check this mass estimate. We derive a total mass of $1.9\times
10^5$~M$_{\odot}$ from the $R$-band flux and the mass-to-light ratio
obtained from the Starburst99 models, which is fully consistent with
the FUV estimate.

If the progenitor star of SN~2008jb was formed in this star-formation
episode 9~Myr ago, then we infer an initial main-sequence mass of
$\simeq 23$~M$_{\odot}$ for the progenitor from the Geneva stellar
evolution models with extended mass-loss used in Starburst99 (Lejeune \&
Schaerer 2001). This estimate is obtained as the maximum main-sequence
mass of a star from an isochrone of 9~Myr of age. We note that this
estimate is based only in the total H$\alpha$ and FUV fluxes of the
region, and a more detailed analysis of the SEDs of the individual
stellar clusters is in preparation.

The equivalent width (EW) of the H$\alpha$ ($EW=263$~\AA) and H$\beta$
($EW=53$~\AA) emission lines in the spectrum of the \ion{H}{2} region
can also be used to estimate an age for the star-forming region using
the Starburst99 models (e.g., Stasi{\'n}ska \& Leitherer 1996; Schaerer
\& Vacca 1998; Leloudas et al. 2011). Essentially, the EW of the
recombination lines gives an estimate of the young over the old stellar
population (see Leitherer 2005, and references therein). We obtain an
age of $6$~Myr from H$\alpha$ EW and $5$~Myr from H$\beta$ EW. These
estimates are $3-4$~Myr ($\sim 3\sigma$ using the statistical
uncertainty) younger than the age derived from the H$\alpha$-to-FUV
ratio of the whole star-forming region. The differences might imply
multiple star-formation episodes within the star-forming region,
however, this would need to be tested with a detailed study of the
stellar populations within the region.   

We can estimate the SFR of the star-forming region using the FUV
luminosity with the calibrations of Lee et al. (2009), finding $\rm
SFR_{UV} =0.01$~M$_{\odot}$~yr$^{-1}$, or about 30\% of the total SFR
of the dwarf galaxy. We note that the H$\alpha$-estimated SFR is a
factor of $\sim 8$ lower than the FUV value, a systematic difference
seen in dwarf galaxies at low SFRs which lacks a satisfactory
explanation (e.g., Lee et al. 2009; see also Fumagalli et al. 2011).

\section{Summary and Conclusions}
\label{sec3}

We have presented the discovery, follow-up observations, and analysis
of SN~2008jb, a bright type~II supernova in the metal-poor, southern
dwarf irregular galaxy ESO~302$-$14 at $\sim 10$~Mpc. This $13.6$~mag
supernova was found in archival data obtained by the CRTS and ASAS
all-sky surveys. This transient was missed by galaxy-targeted
supernova surveys like CHASE and by amateur astronomers mainly because
the host galaxy is a low-luminosity dwarf, $\sim 1$~mag fainter than
the SMC, and targeted surveys use catalogs that are incomplete for
small galaxies (e.g., Leaman et al. 2011).

SN~2008jb has $V$- and $I$-band light curves similar to normal
type~IIP supernovae, with peak magnitude $M_V \simeq -16.5$~mag, but
it can also be classified as an intermediate case between type~IIP and
type~IIL due to its faster initial decay, perhaps similar to SN~1999ga
(Pastorello et al. 2009b). It shows a $\sim 95$~day plateau and a fast
$\sim 1.4$~mag decline to a late-time decline slope of $0.013$~mag/day
in the $V$-band. This decline is consistent with the radioactive decay
of $^{56}$Co to $^{56}$Fe, and argues for $0.04 \pm 0.01$~M$_{\odot}$
of $^{56}$Ni synthesized in the explosion that is powering the
lightcurve. We detect mid-IR emission from SN~2008jb $8-14$ months
after explosion in three epochs of archival Spitzer and WISE data. The
mid-IR emission has an SED with black-body temperature of $600-700$~K,
characteristic of the warm dust emission seen in some nearby
core-collapse supernovae and luminous transients (e.g., Kotak et
al. 2009; Prieto et al. 2009). The evolution of the mid-IR emission
with time is consistent with the products of radioactive decay heating
the dust. The characteristic mid-IR dust radius shrinks with time, an
evolution that is not typically seen in normal type~IIs (some
exceptions are SN~2007it and SN~2007od; Andrews et al. 2010, 2011).

We obtained a spectrum of SN~2008jb about 2 years after the explosion.
It displays a very broad (${\rm FWHM}\simeq 14000$~km~s$^{-1}$), boxy,
and flat-topped H$\alpha$ emission line, leading to its type~II
supernova spectroscopic classification. The broad and boxy line
profile seen in H$\alpha$ is quite unusual for normal type~IIP
supernovae at late times, but has been seen in some objects like
SN~1993J, SN~2007od, and also a few well-studied type~IIL
supernovae. We find that the H$\alpha$ line luminosity is in excess of
the expected luminosity from radioactive $^{56}$Co decay predicted by
the models of Chugai (1991) for total ejected masses $\rm M=5, 14,
20$~M$_{\odot}$. This indicates that there is an external source of
energy, like ejecta-CSM interaction, and/or the mass of the progenitor
star is $\gtrsim 20$~M$_{\odot}$. We do not see clear signs of
ejecta-CSM interaction like narrow lines or irregularities in the
H$\alpha$ profile. It would be interesting to obtain late time X-ray
and radio observations in order to have independent constraints on the
importance of ejecta-CSM interaction in SN~2008jb.

We studied the host galaxy environment of SN~2008jb in ESO~302$-$14 with
optical spectra and imaging. Using the spectrum of an \ion{H}{2} region
at $\sim 150$~pc from the supernova site we measure an oxygen abundance
of $\rm 12+log(O/H) = 8.21 \pm 0.03$ (PP04~$O3N2$ method) and $\rm
12+log(O/H) = 8.13 \pm 0.03$ (PP04~$N2$ method) from the strong nebular
emission lines, which is similar to the SMC and one of the lowest
measured metallicities of local core-collapse supernovae environments
(e.g., in the lower 3\% of measured oxygen abundances of the
core-collapse sample of Anderson et al. 2010). The supernova exploded in
a large star-forming complex with strong optical and GALEX FUV emission
which is surrounded by a large H$\alpha$ ring with $\rm R \simeq
350$~pc. The H$\alpha$-to-FUV ratio of this region is consistent with a
stellar population with an age of $\sim 9$~Myr derived from Starburst99
modeling and single-age stellar population models. This age implies a
supernova progenitor mass of $\simeq 23$~M$_{\odot}$, assuming a single
star (but see, e.g., Smith et al. 2011a for the importance of binary
progenitors), if the progenitor formed in this star-forming episode. The
equivalent width of the H$\alpha$ and H$\beta$ emission lines in the
spectrum of the \ion{H}{2} region give another constraint on the age of
the region of $5-6$~Myr, $3-4$~Myr younger than the estimate from
H$\alpha$-to-FUV ratio. We note that the star-formation history could be
more complicated than a single burst, and we plan to study the region in
detail using multiwavelength data in a future study. In particular, it
would be interesting to include high-resolution data from HST.

Large expanding H$\alpha$ shells (supershells with radii $>300$~pc) have
been observed and studied in many nearby star-forming dwarf galaxies and
have typical dynamical ages of $\sim 10$~Myr (e.g., Martin 1998), which
is fairly consistent with the age we derive here from the H$\alpha$ and
FUV emission. These structures are produced by the combined effect of
many supernova explosions and winds from massive stars (e.g.,
Chakraborti \& Ray 2011), and are the likely precursors of galactic
winds. In a sense, we may perhaps be witnessing supernova feedback in
real-time in this star-forming region. 

The bias to large star-forming galaxies is clearly present in the
samples of nearby \mbox{($\lesssim 30$~Mpc)} core-collapse supernovae
used for progenitor studies (e.g., Smartt et al. 2009), and has been
discussed in detail as a possible explanation for the discrepancy
between measured local supernova rates and predicted supernova rates
from galaxy star-formation rates (e.g., Horiuchi et al. 2011). Since
the environments of nearby core-collapse supernovae used for
progenitor studies generally miss dwarf galaxies because they were not
included in the original searches, the progenitor properties and
conclusions drawn from these samples regarding stellar evolution are
not complete. In particular, the dearth of high-mass ($\sim
20-30$~M$_{\odot}$) progenitor stars of type~IIP supernovae (``red
supergiant problem") could be alleviated if these progenitor stars
prefer lower-metallicity environments. For example, this could be
caused by evironmental variations in the stellar IMF (e.g., Meurer et
al. 2009) or by changes in the fraction of type~II spectroscopic
subtypes as a function of metallicity due to stellar evolution (e.g.,
Arcavi et al. 2010). Indeed, we find that the properties of the
spectrum of SN~2008jb are more consistent with a massive progenitor
(but see discussion about supernova modeling in, e.g., Smartt et al.
2009; Bersten \& Hamuy 2009; Bersten et al. 2011), and the
star-forming region where it was found has a young age compared with
the ages of detected type~IIP progenitors ($\gtrsim 15$~Myr).

Interestingly, a strong preference for low-metallicity hosts is
observed in long GRBs (e.g., Stanek et al. 2006) and luminous
core-collapse supernovae (e.g., Neill et al. 2011; Stoll et al. 2011),
which have been linked with massive star progenitors ($\rm M \gtrsim
20-30$~M$_{\odot}$). SN~2008jb offers the unique chance of studying in
detail a nearby type~II supernova with host properties similar to long
GRBs and the most luminous core-collapse supernovae.

The mapping between different classes of massive stars and their
supernovae is not yet fully understood. Special insights are expected
to be obtained when unusual explosions can be connected to unusual
progenitor stars and galaxy hosts. Nearby objects are especially
useful in terms of larger fluxes for an extended time after explosion,
better spatial resolution for progenitor studies, and improved
prospects for detection by new messengers like gamma rays (e.g.,
Timmes \& Woosley 1997; Horiuchi \& Beacom 2010), neutrinos (e.g.,
Ando \& Beacom 2005; Kistler et al. 2011), and gravitational waves
(e.g., Ott 2009). In addition, data from these objects are needed for
a comprehensive understanding of the nearby universe.

It is difficult to find the nearest supernovae in small host galaxies
with searches that target individual (generally large) galaxies, like
LOSS, CHASE, and also amateur efforts. And it is difficult to find
them with volume-based searches such as PTF (Rau et al. 2009) and
Pan-STARRS (Kaiser et al. 2002) that have a deep, but relatively small
survey area with good cadence. A shallower all-sky survey with
excellent cadence, like ASAS, will help us find nearby
\mbox{($\lesssim 30$~Mpc)} supernovae in all kinds of environments,
including low-metallicity dwarf galaxies like the host of SN~2008jb
(see also Khan et al. 2011, Stoll et al. 2011, for other supernovae
studied with ASAS). Upgrades that will significantly increase the
sensitivity and response speed of ASAS are underway.

\acknowledgments

We thank John~Mulchaey for obtaining one of the Magellan images
presented in this work, Chris~Kochanek for detailed comments,
Rupali~Chandar, Crystal~Martin, and Linda~Watson for discussions, and
Chris~Burns for providing his image subtraction code. We also thank the
anonymous referee for a careful reading of the manuscript. We are
indebted to the staff of Las Campanas Observatory for their assistance.
JLP acknowledges support from NASA through Hubble Fellowship grant
HF-51261.01-A awarded by STScI, which is operated by AURA, Inc. for
NASA, under contract NAS~5-2655. JFB is supported by the National
Science Foundation CAREER Grant PHY-0547102. GP is supported by the
Polish MNiSW grant N203 007 31/1328. KSZ and DMS are supported in part
by NSF grant AST-0908816. Support for MC is provided by the Ministry for
the Economy, Development, and Tourism's Programa Iniciativa Cient\'ifica
Milenio through grant P07-021-F, awarded to The Milky Way Millenium
Nucleus; by Proyecto Basal PFB-06/2007; by FONDAP Centro de
Astrof\'isica 15010003; and by proyecto FONDECYT Regular \#1110326. The
CRTS is supported in part by the NSF grant AST-0909182. This research
has made extensive use of the NASA/IPAC Extragalactic Database (NED)
which is operated by the JPL, Caltech, under contract with NASA.

\newpage

\begin{figure}[t]
\plotone{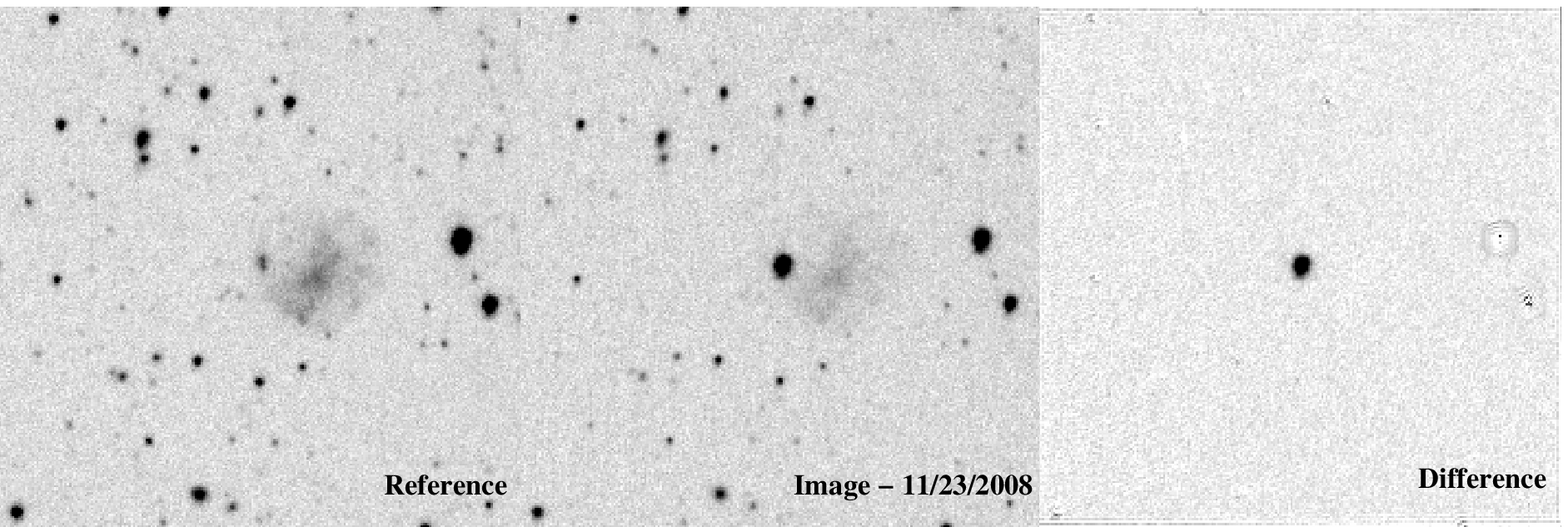}
\caption{Siding Spring Survey images ($3\farcm 6\times 3\farcm 6$)
  centered on SN~2008jb. {\it Left panel}: reference image obtained
  from a combination of pre-discovery images. {\it Middle panel}:
  first SSS post-discovery image obtained on Nov. 23, 2008. {\it Right
    panel}: difference image. In all the panels, North is up and East
  is to the left.}
\label{fig:diffim}
\end{figure}

\begin{figure}[t]
\plotone{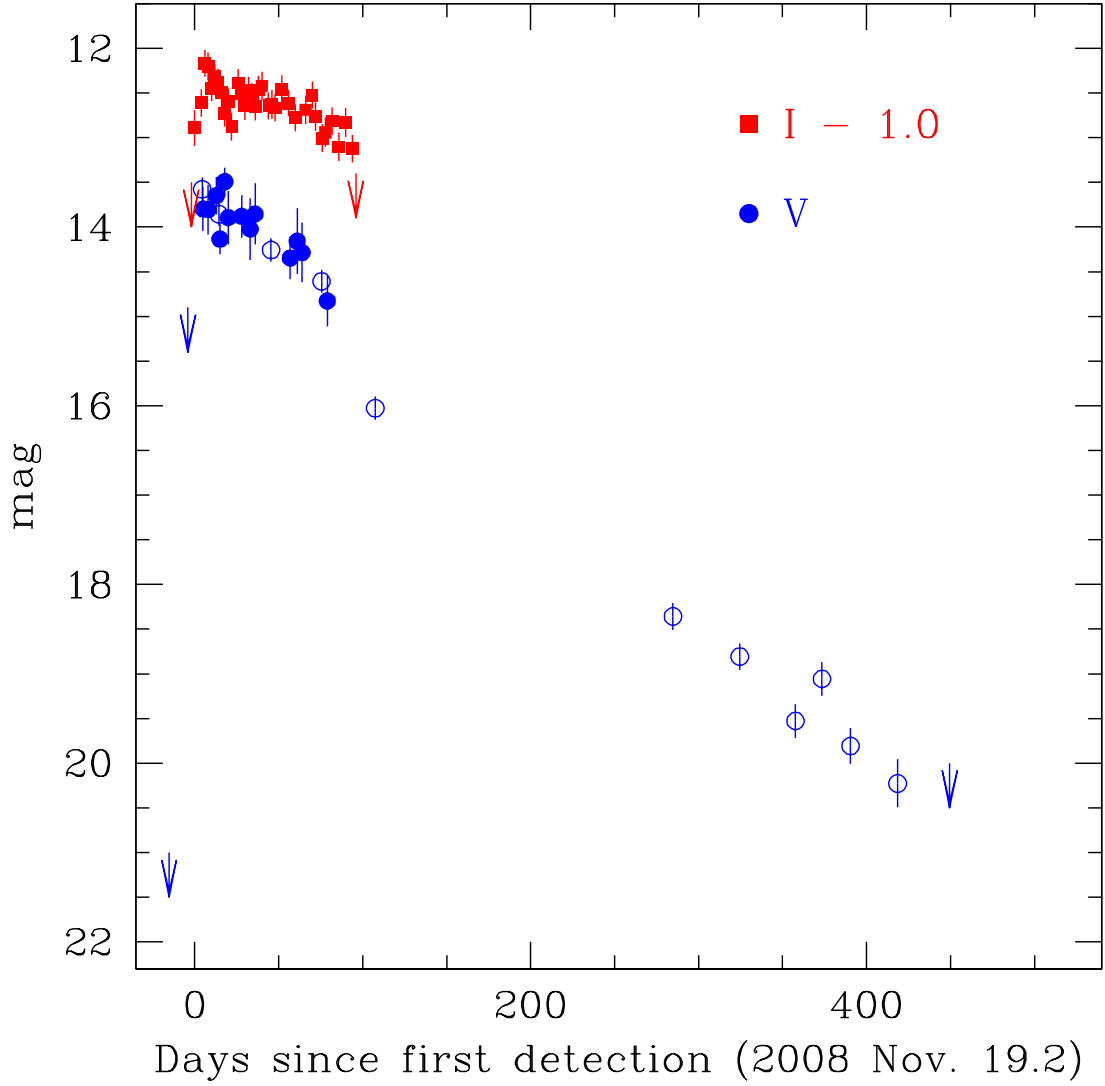}
\caption{Light curves of SN~2008jb obtained from CRTS (open symbols;
  unfiltered, calibrated in the $V$-band) and ASAS (filled symbols;
  $V$ and $I$-band) photometric data. The arrows are 3$\sigma$ upper
  limits on the magnitudes.}
\label{fig:lc}
\end{figure}

\begin{figure}[t]
\plotone{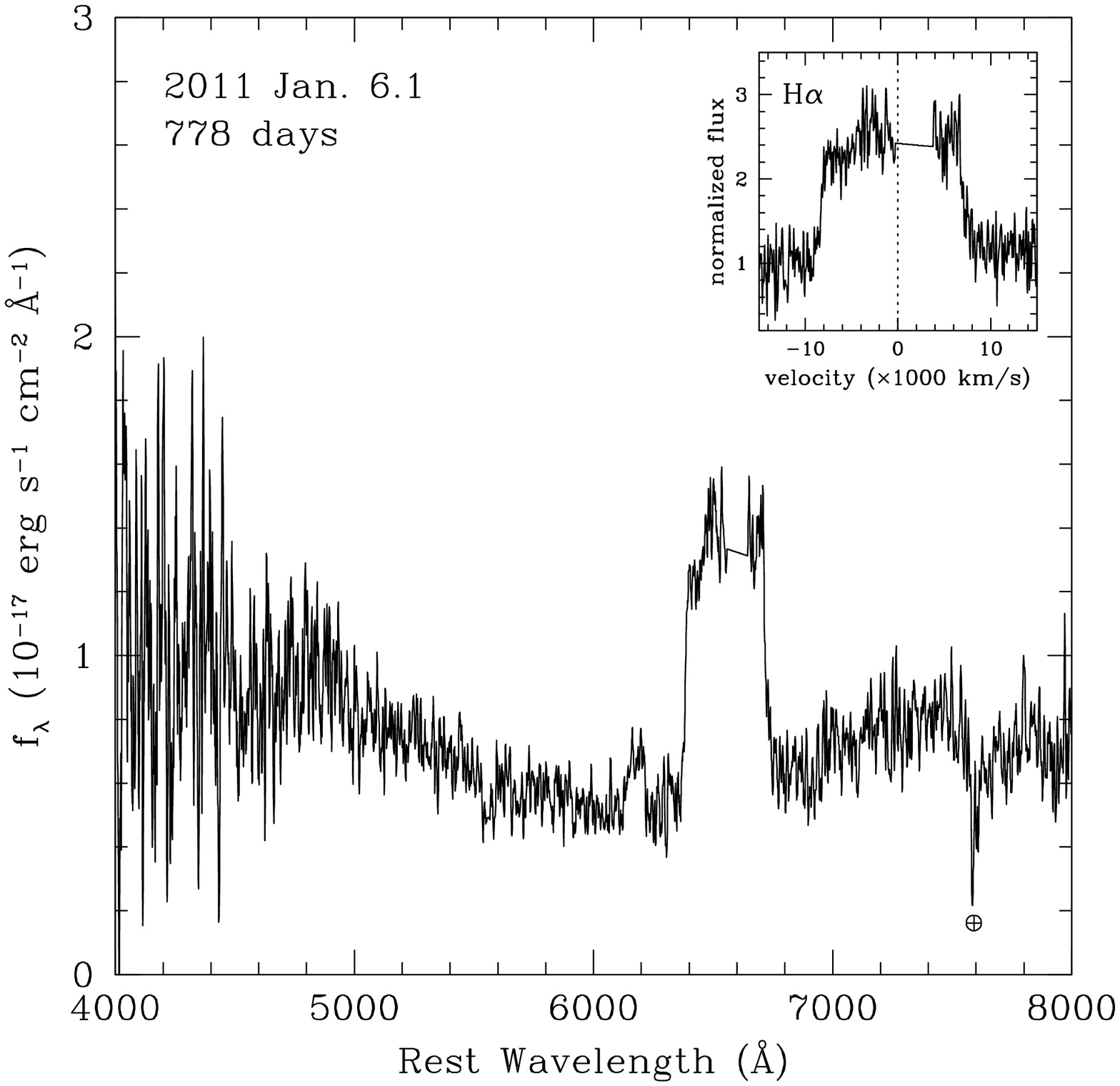}
\caption{Late-time optical spectrum of SN~2008jb obtained with
  Magellan/IMACS on 2011 Jan. 6.1, $778$ days after the first
  detection of the supernova with ASAS. The wavelength has been
  corrected to the rest frame using the recession velocity of the host
  galaxy \mbox{($v=872$~km~s$^{-1}$)}. The spectrum is dominated by a
  broad, flat-topped H$\alpha$ emission line \mbox{(${\rm FWHM}\simeq
    14000$~km~s$^{-1}$)} shown in the inset panel. We have linearly
  interpolated the fluxes in the chip gap between $6588$ and
  6662~\AA\ (observed wavelength). The fluxes have been corrected for
  Galactic extinction and smoothed using a 5-pixel boxcar.}
\label{fig:snspec}
\end{figure}

\begin{figure}[t]
\plotone{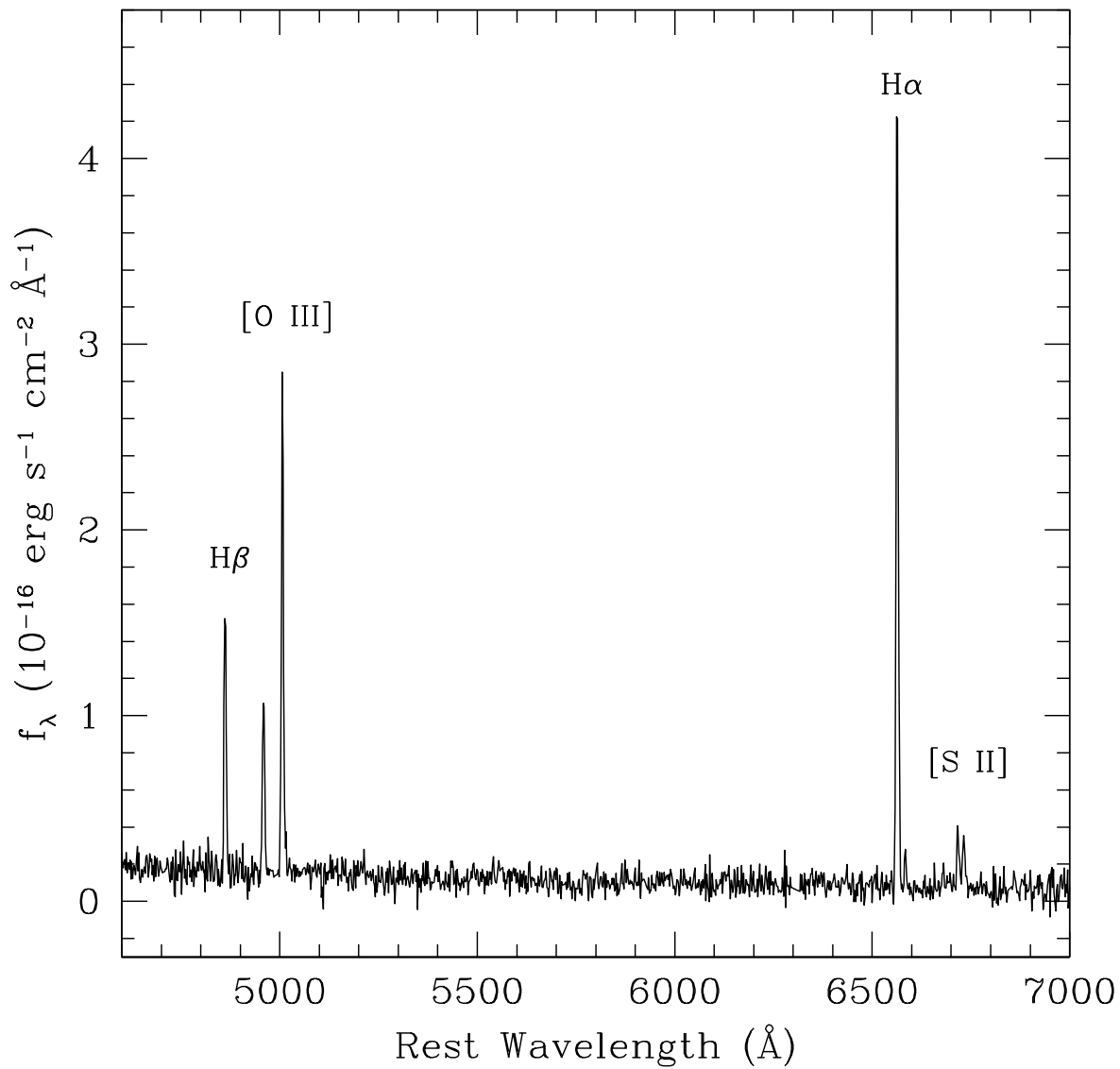}
\caption{Optical spectrum obtained with du\,Pont/WFCCD of a nearby
  \ion{H}{2} region in the dwarf irregular host galaxy of SN~2008jb,
  ESO~302$-$14. The spectrum shows Balmer recombination lines
  (H$\alpha$ and H$\beta$) and forbidden emission lines ([\ion{O}{3}]
  $\lambda\lambda$~4959, 5007, [\ion{S}{2}] $\lambda\lambda$~6713,
  6731, and weak \mbox{[\ion{N}{2}] $\lambda$~6583}). The wavelength
  has been corrected to the rest-frame using the recession velocity of
  the host galaxy. The fluxes have been corrected for Galactic
  extinction.}
\label{fig:galspec}
\end{figure}

\begin{figure}[t]
\plotone{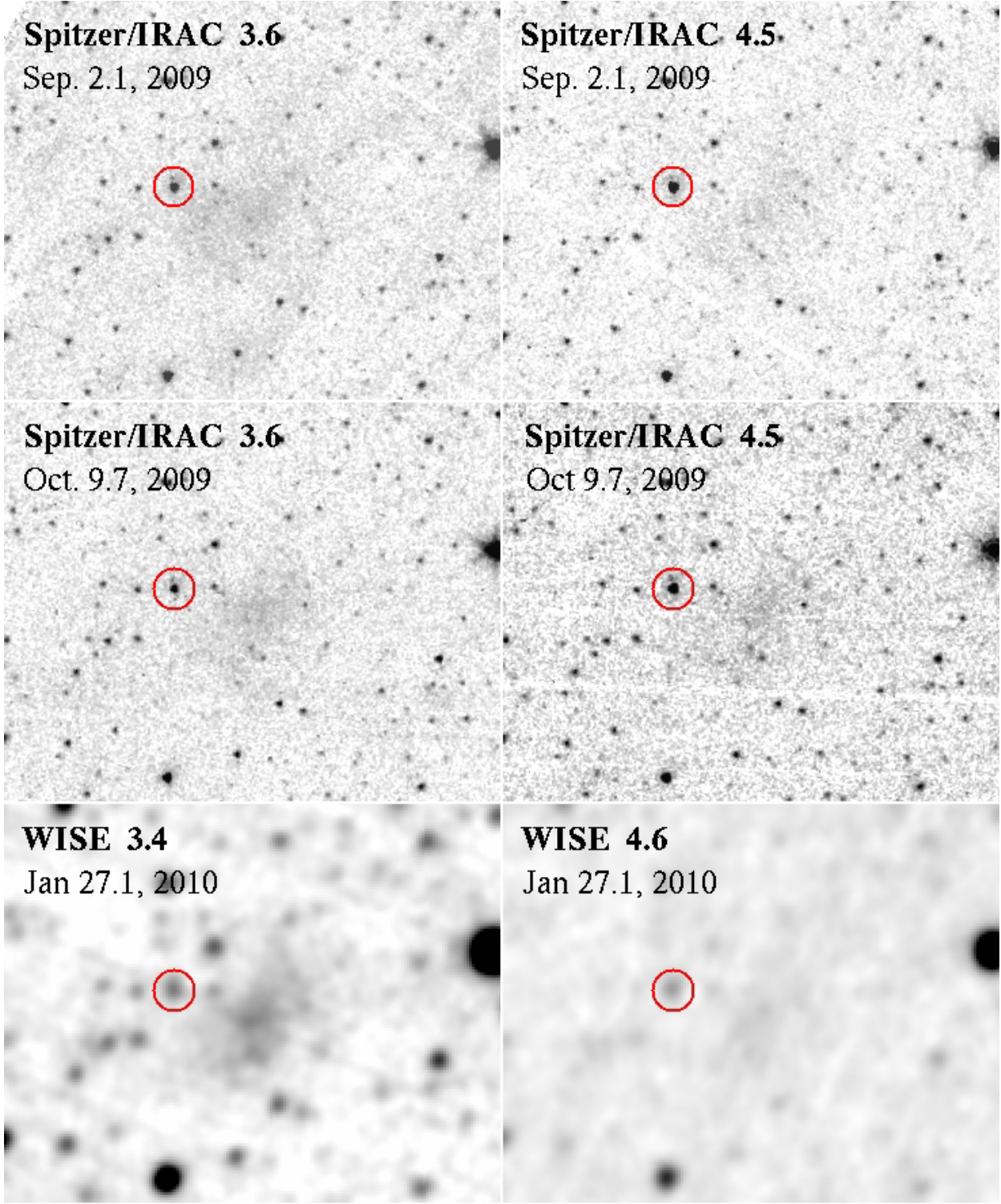}
\caption{Spitzer IRAC (3.6~$\mu$m, 4.5~$\mu$m) and WISE (3.4~$\mu$m
  and 4.6~$\mu$m) images showing the late-time detections of SN~2008jb
  (red circle) at three different epochs. Each $4\farcm 1 \times
  3\farcm 3$ panel is centered on the host galaxy of SN~2008jb. In
  all panels, North is up and East is to the left.}
\label{fig:ir}
\end{figure}

\begin{figure}[t]
\plotone{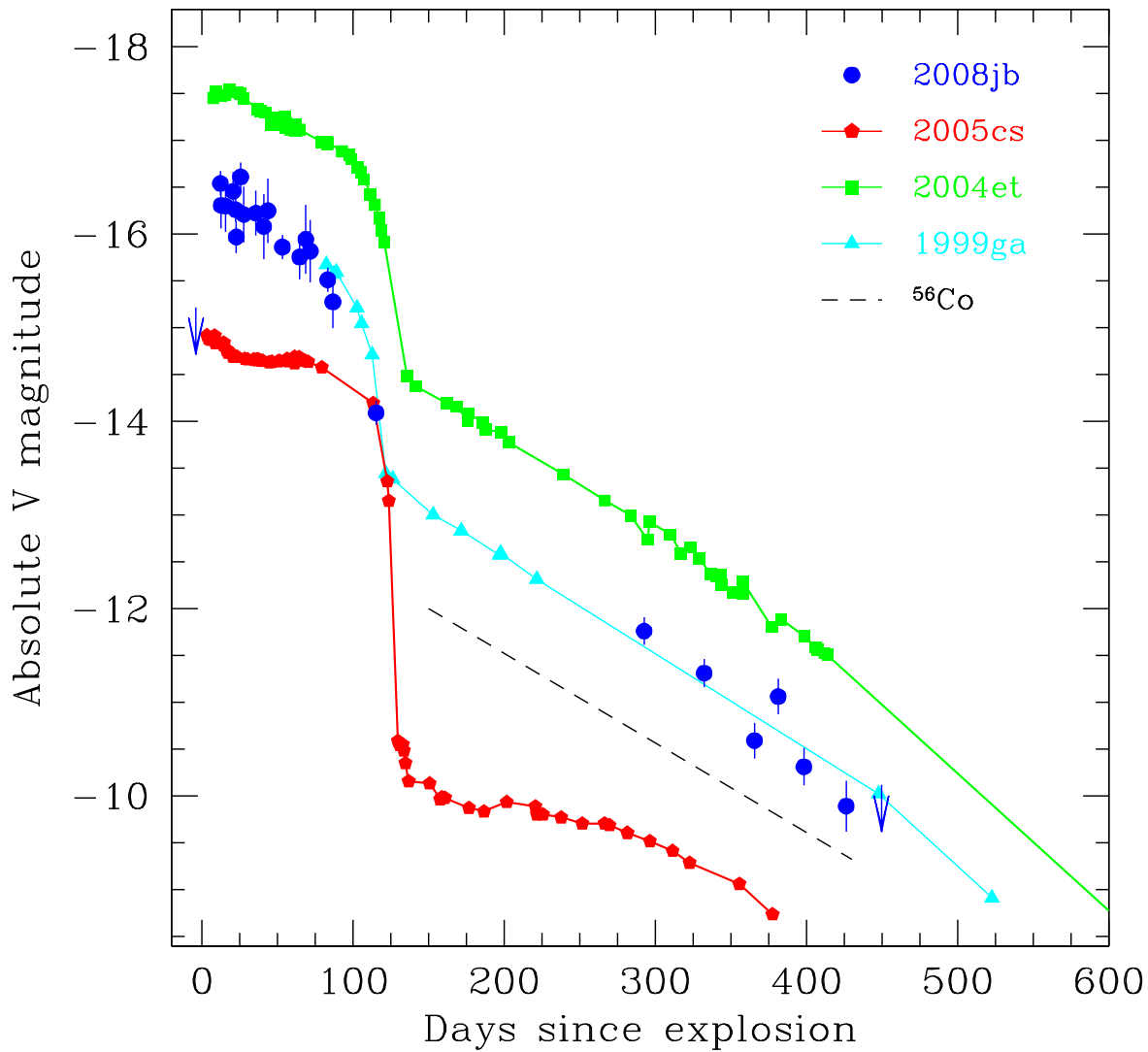}
\caption{Absolute magnitude light curves in $V$-band of type~II
  supernovae compared with SN~2008jb (filled circles). The comparison
  light curves are: the low-luminosity type~IIP 2005cs (filled
  pentagons; Pastorello et al. 2009a), the luminous type~IIP 2004et
  (filled squares; Maguire et al. 2010), and the low-luminosity type
  IIL 1999ga (filled triangles; Pastorello et al. 2009b). The dashed
  line shows the slope of the $^{56}$Co decay
  ($0.01$~mag~day$^{-1}$). The magnitudes have been corrected by
  Galactic and internal extinction in each case.}
\label{fig:lcabs}
\end{figure}

\begin{figure}[t]
\plotone{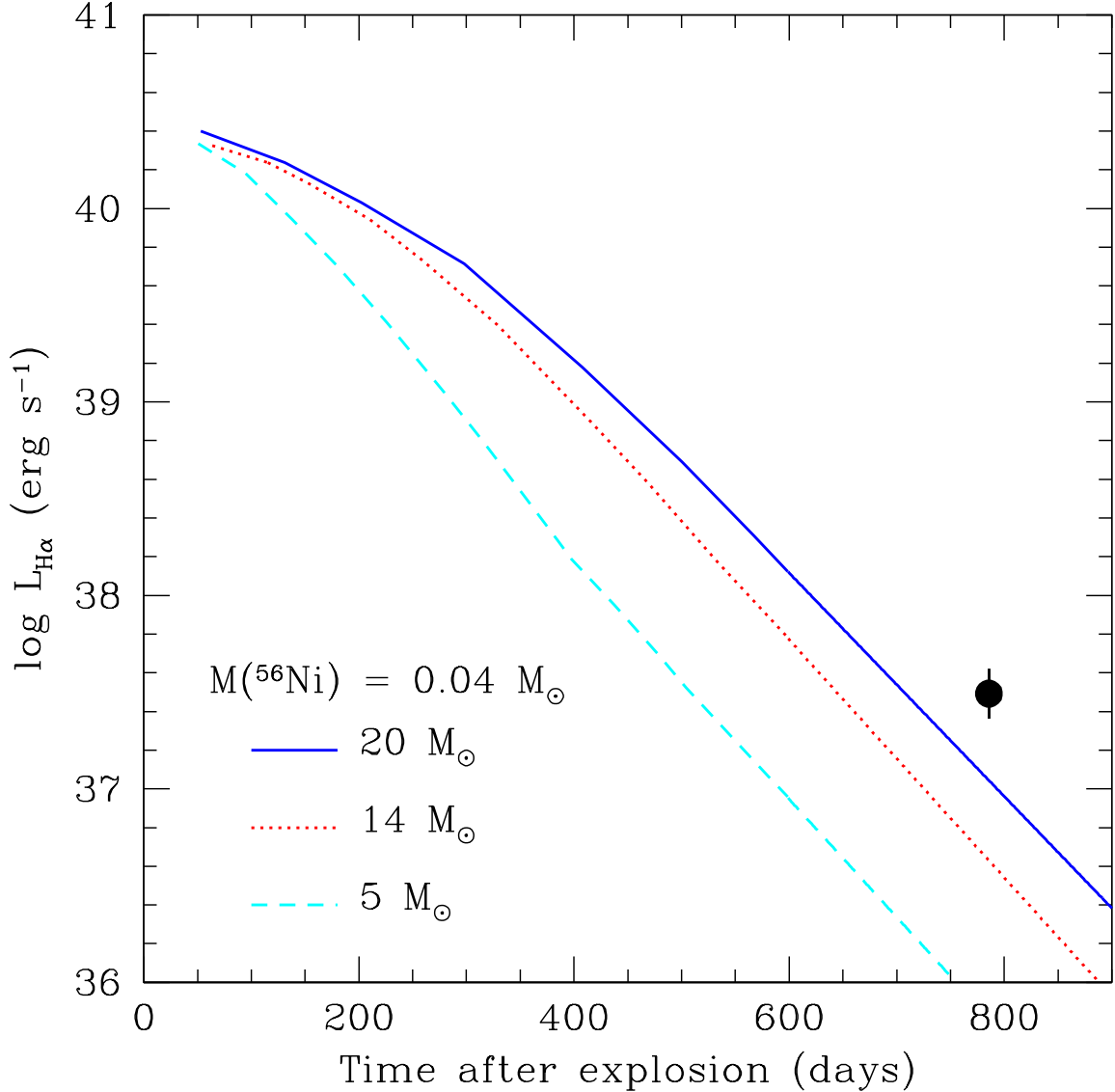}
\caption{Evolution of the H$\alpha$ emission line luminosity as a
  function of time from the models of Chugai (1991) and taken from the
  parametrization presented in Pastorello et al. (2009b). The filled
  circle is the measurement from the spectrum of SN~2008jb assuming a
  conservative error of 30\% (which includes calibration) in
  luminosity. These models trace the evolution of the H$\alpha$ line
  luminosity under the assumption of pure radioactive decay of
  $^{56}$Co. The different lines correspond to ejected masses of
  20~M$_{\odot}$ (continuous line), 14~M$_{\odot}$ (dotted line), and
  5~M$_{\odot}$ (dashed line). Here we have scaled the model curves to
  a $^{56}$Ni mass of 0.04~M$_{\odot}$ consistent with SN~2008jb.}
\label{fig:halpha}
\end{figure}

\begin{figure}[t]
\plotone{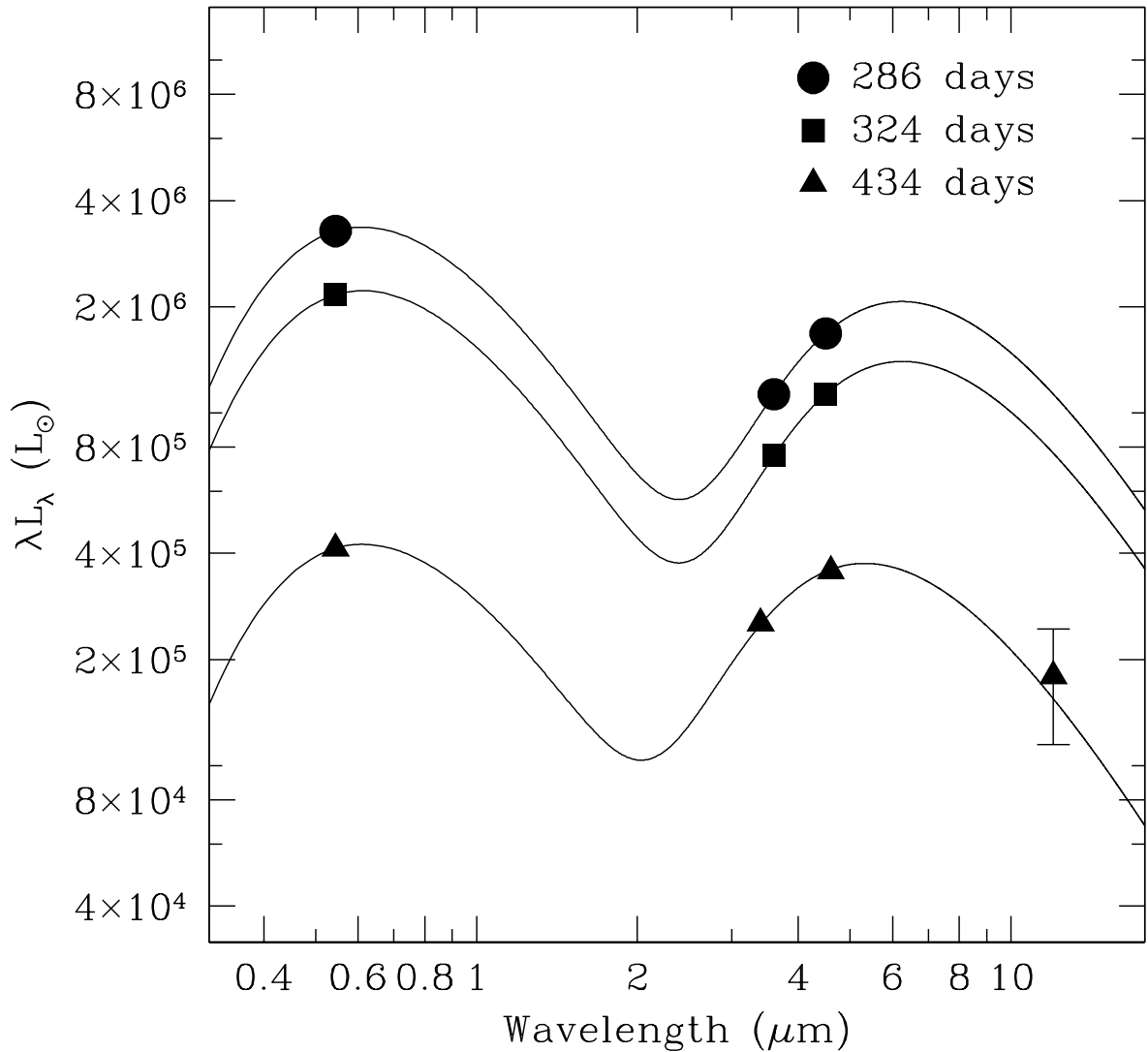}
\caption{Spectral energy distribution of SN~2008jb at three late-time
  epochs including $V$-band, Spitzer IRAC 3.6~$\mu$m and 4.5~$\mu$m,
  and WISE 3.4~$\mu$m, 4.6~$\mu$m, and 12~$\mu$m fluxes. The fluxes
  have been corrected for Galactic and host extinction. The lines are
  fits to the data using the sum of two blackbodies. For the ``hot''
  blackbody component we assume $T_{\rm hot}=6000$~K in all three
  epochs, and fit for the absolute normalization. The ``warm''
  component is also a blackbody, but we fit for the absolute
  normalization and temperature. The temperature of the warm component
  is between $\sim 600-700$~K in the three epochs. The labels show the
  epochs (in days) after the first detection.}
\label{fig:sedir}
\end{figure}

\begin{figure}[t]
\plotone{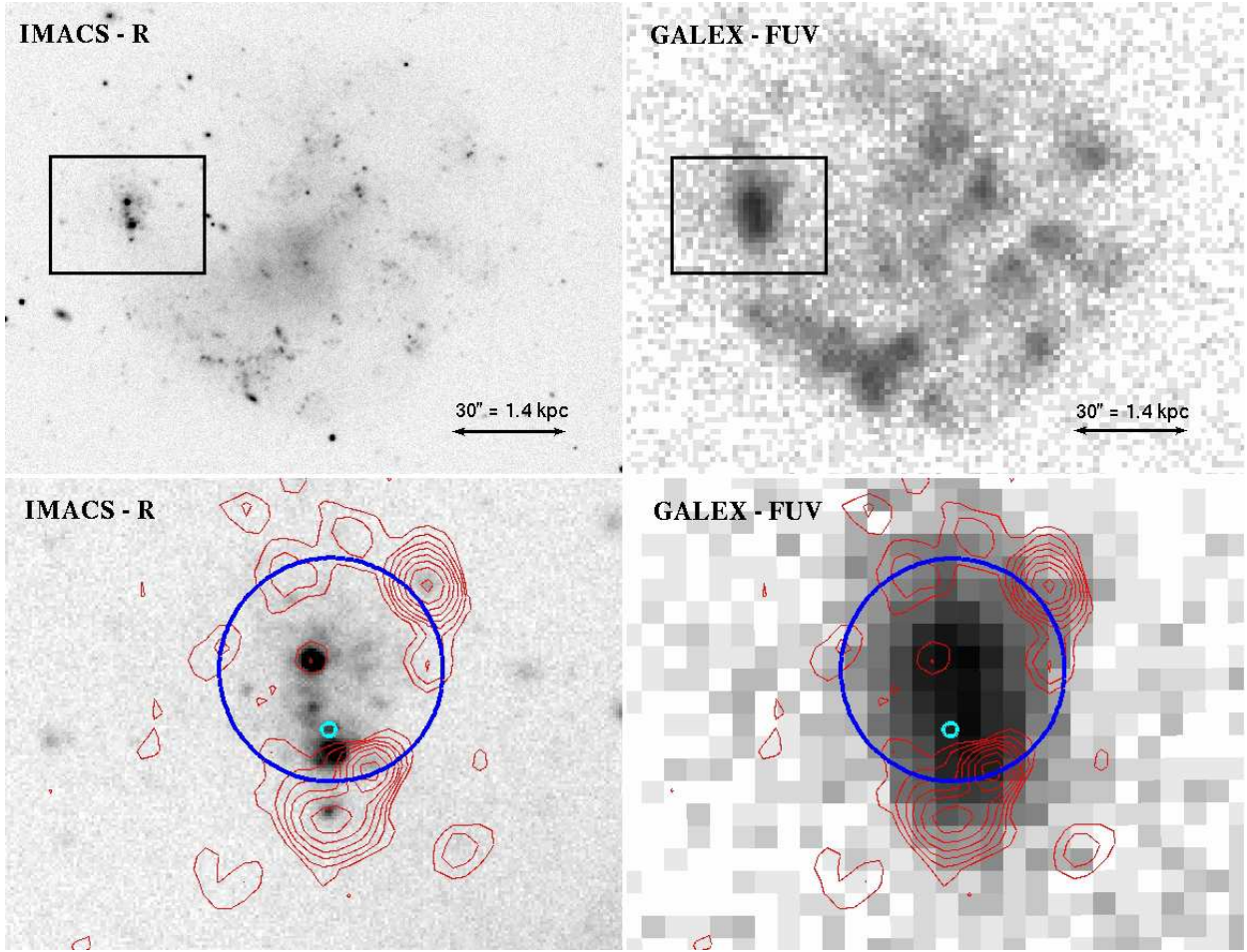}
\caption{Host galaxy environment of SN~2008jb. {\it Top panels:}
  Magellan/IMACS $R$-band image obtained on 2010 Nov. 10 ({\it left})
  and GALEX/FUV ($\lambda_{\rm eff}=1539$~\AA) archival image obtained
  on 2004 Nov. 18 ({\it right}). The FWHM of stars in the images are
  0$\farcs$7 ($R$-band) and $4\arcsec$ (FUV). {\it Bottom panels:}
  Zoomed-in versions of the rectangular regions around the supernova
  position shown in the top panels. The red contours trace H$\alpha$
  emission from archival images obtained by the SINGG survey on 2000
  Oct. 28. The blue circle has a diameter of $15'' \approx 700$~pc and
  approximately follows the H$\alpha$ emission around the
  star-formation complex. The cyan circle shows the position of
  SN~2008jb. In all panels, North is up and East is to the left.}
\label{fig:gal}
\end{figure}

\begin{figure}[t]
\plotone{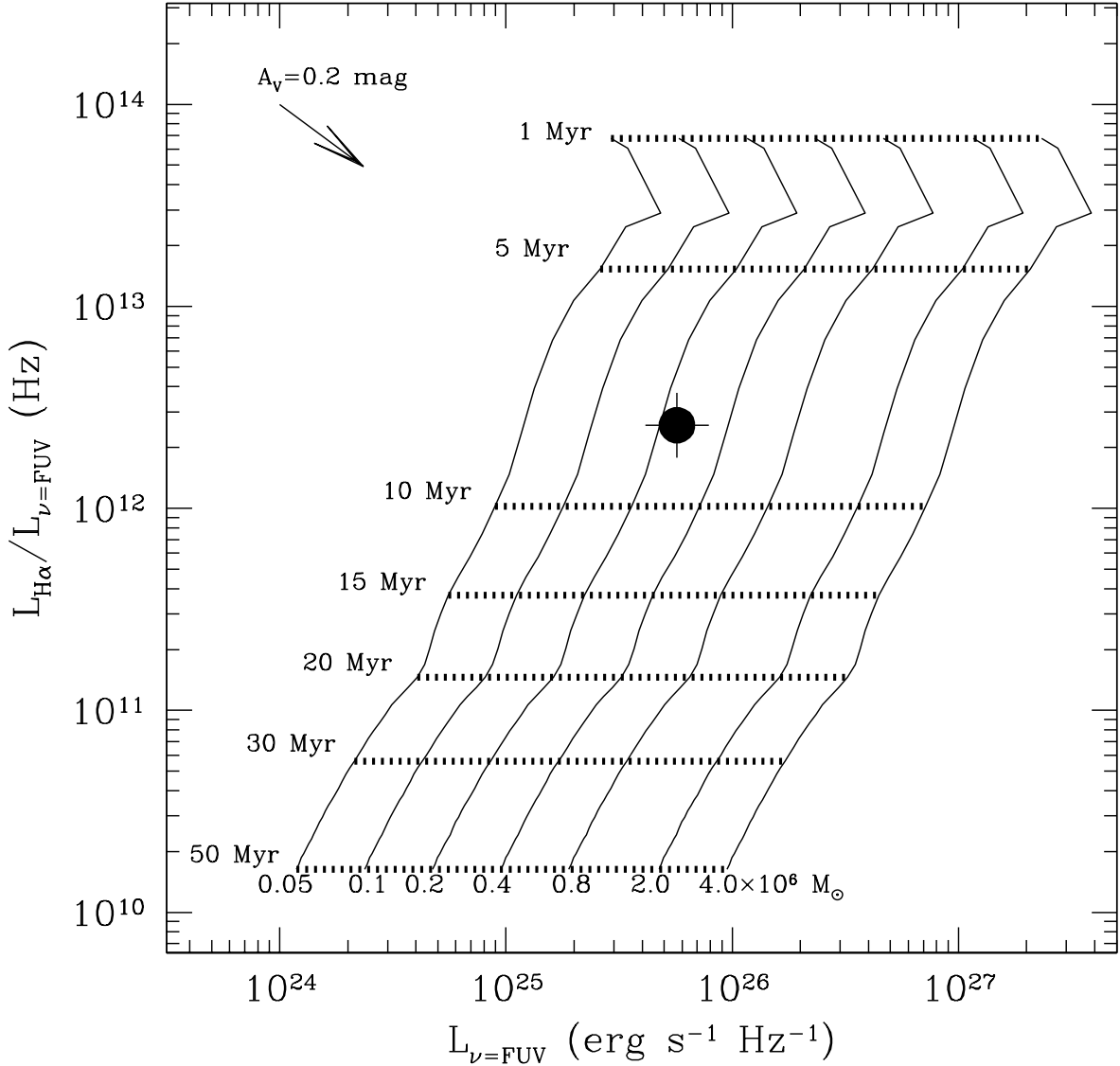}
\caption{H$\alpha$/FUV ratio as a function of the specific FUV
  luminosity. The filled circle shows the integrated value for the
  star-formation complex that hosts SN~2008jb in ESO~302$-$14, after
  correcting for the total extinction. The lines show theoretical
  Starburst99 models for single-burst clusters with masses $5\times
  10^4 - 4\times 10^{6}$~M$_\odot$ and ages in the range $1-50$~Myr,
  assuming a metallicity of 0.2~Z$_{\odot}$ and a Salpeter IMF. If the
  complex that hosts SN~2008jb is from a single burst of
  star-formation, it is consistent with having a total mass \mbox{$\rm
    M= (2.4\pm 1.0)\times 10^{5}$~M$_{\odot}$} and age \mbox{$8.8 \pm
    0.8$~Myr}. The vector in the top left shows the effect of $A_V =
  0.2$~mag of host extinction in the fluxes.}
\label{fig:sfregion}
\end{figure}

\clearpage

\begin{deluxetable}{lcc}
\tablecolumns{3}
\tablewidth{0pt}
\tablecaption{CRTS $V$-band lightcurve of SN~2008jb}
\tablehead{
\colhead{HJD - 2450000} &
\colhead{mag} & 
\colhead{$\sigma$} \\
\colhead{(day)} &
\colhead{} & 
\colhead{} }
\startdata
4774.21 & $> 21.0$ & \\
4794.12 & 13.58 & 0.08 \\ 
4804.14 & 13.86 & 0.08 \\ 
4835.09 & 14.26 & 0.08 \\ 
4864.96 & 14.61 & 0.08 \\ 
4896.96 & 16.03 & 0.08 \\ 
5074.28 & 18.36 & 0.09 \\ 
5114.17 & 18.81 & 0.09 \\ 
5147.21 & 19.53 & 0.11 \\ 
5163.15 & 19.06 & 0.11 \\ 
5180.08 & 19.81 & 0.12 \\ 
5208.05 & 20.23 & 0.16 \\
5238.93 & $>20.0$ & \\
\enddata
\end{deluxetable}

\begin{deluxetable}{lcc}
\tablecolumns{3} \tablewidth{0pt} \tablecaption{ASAS lightcurve of
  SN~2008jb} \tablehead{ \colhead{HJD - 2450000} & \colhead{mag} &
  \colhead{$\sigma$} \\ \colhead{(day)} & \colhead{} & \colhead{} }
\startdata
% V-band data
\hline
\multicolumn{3}{c}{$V$} \\
\hline
4785.66 &   $> 14.9$ & \\
4794.68 & 13.80 & 0.25 \\  
4797.69 & 13.80 & 0.28 \\  
4802.64 & 13.65 & 0.21 \\  
4804.73 & 14.13 & 0.17 \\  
4807.59 & 13.49 & 0.16 \\  
4809.65 & 13.89 & 0.30 \\  
4817.64 & 13.88 & 0.24 \\  
4822.77 & 14.02 & 0.35 \\  
4825.65 & 13.85 & 0.34 \\  
4846.61 & 14.35 & 0.24 \\  
4850.73 & 14.16 & 0.37 \\  
4853.64 & 14.29 & 0.33 \\  
4868.60 & 14.83 & 0.28 \\
% I-band data
\hline
\multicolumn{3}{c}{$I$} \\
\hline
4787.65 & $> 14.5$ & \\
4789.70 & 13.89 & 0.20 \\  
4793.70 & 13.61 & 0.15 \\  
4795.70 & 13.17 & 0.15 \\  
4797.71 & 13.20 & 0.16 \\  
4799.72 & 13.45 & 0.15 \\  
4801.68 & 13.32 & 0.15 \\  
4803.65 & 13.38 & 0.16 \\  
4805.65 & 13.50 & 0.16 \\  
4807.64 & 13.72 & 0.15 \\  
4809.65 & 13.60 & 0.15 \\  
4811.68 & 13.88 & 0.16 \\  
4815.67 & 13.39 & 0.16 \\  
4817.67 & 13.51 & 0.15 \\  
4819.67 & 13.65 & 0.15 \\  
4821.66 & 13.47 & 0.15 \\  
4823.66 & 13.54 & 0.15 \\  
4825.66 & 13.65 & 0.16 \\  
4827.67 & 13.46 & 0.15 \\  
4829.68 & 13.42 & 0.16 \\  
4833.64 & 13.64 & 0.15 \\  
4835.62 & 13.63 & 0.16 \\  
4837.62 & 13.66 & 0.15 \\  
4841.64 & 13.46 & 0.15 \\  
4845.64 & 13.61 & 0.15 \\  
4849.63 & 13.78 & 0.15 \\  
4855.63 & 13.69 & 0.16 \\  
4859.62 & 13.52 & 0.15 \\  
4861.62 & 13.76 & 0.16 \\  
4865.61 & 14.00 & 0.15 \\  
4867.60 & 13.94 & 0.17 \\  
4871.61 & 13.81 & 0.15 \\  
4875.61 & 14.10 & 0.16 \\  
4879.60 & 13.83 & 0.16 \\  
4883.59 & 14.12 & 0.15 \\ 
4885.60 & $> 14.4$ & \\
\enddata
\end{deluxetable}

%%%%%%%%%%%%%%%%%%%%%%%%%%%%%%%%%%%%%%%%%

\begin{deluxetable}{lcl}
\tablecolumns{2}
\tablewidth{0pt}
\tablecaption{Emission line fluxes of \ion{H}{2} region in ESO~302$-$14}
\tablehead{
\colhead{Line} & 
\colhead{Flux$^{\rm a}$ (10$^{-15}$ erg~s$^{-1}$~cm$^{-2}$)}}
\startdata
H$\beta$ & $0.94 \pm 0.09$ \\
$[$\ion{O}{3}$]$~4959 & $0.67\pm 0.07$ \\
$[$\ion{O}{3}$]$~5007 & $1.68\pm0.15$ \\
H$\alpha$ & $2.84\pm 0.21$ \\
$[$\ion{N}{2}$]$~6583 & $0.12 \pm 0.02$ \\
$[$\ion{S}{2}$]$~6713 & $0.22 \pm 0.03$ \\
$[$\ion{S}{2}$]$~6731 & $0.22 \pm 0.03$ \\
\enddata
\tablenotetext{a}{Fluxes have been corrected by Galactic $E(B-V)_{\rm MW}=0.009$~mag using CCM reddening law.}
\end{deluxetable}

%%%%%%%%%%%%%%%%%%%%%%%%%%%%%%%%%%%%%%%%

\begin{deluxetable}{lcccc}
\tablewidth{0pt}
\tablecaption{Spitzer and WISE photometry of SN~2008jb}
\tablehead{
\colhead{HJD} & 
\colhead{Band} & 
\colhead{Flux (mJy)} & 
\colhead{Vega mag} &
\colhead{Instrument}}
\startdata
2455076.12 & 3.6~$\mu$m  &  $0.47\pm0.01$ & $14.43\pm 0.03$ & Spitzer/IRAC\\
2455076.12 & 4.5~$\mu$m  &  $0.88\pm0.03$ & $13.28\pm 0.03$ & Spitzer/IRAC\\
2455113.71 & 3.6~$\mu$m  &  $0.32\pm0.01$ & $14.87\pm 0.03$ & Spitzer/IRAC\\
2455113.71 & 4.5~$\mu$m  &  $0.59\pm0.02$ & $13.71\pm 0.03$ & Spitzer/IRAC\\
2455223.63 & 3.4~$\mu$m  &  $0.10\pm0.01$ & $16.20\pm 0.07$ & WISE \\   
2455223.63 & 4.6~$\mu$m  &  $0.19\pm0.01$ & $14.88\pm 0.07$ & WISE \\ 
2455223.63 & 12~$\mu$m  &  $0.25\pm0.09$ & $12.75\pm 0.40$ & WISE \\   
2455223.63 & 22~$\mu$m  &  $< 1.8$ & $> 9.18$ & WISE \\ 
\enddata
\tablecomments{Spitzer data is from warm Spitzer IRAC program 61060 (PI:~K.~Sheth)}
\end{deluxetable}

%%%%%%%%%%%%%%%%%%%%%%%%%%%%%%%%%%%%%%

\begin{deluxetable}{lcl}
\tablecolumns{3}
\tablewidth{0pt}
\tablecaption{Properties of SN~2008jb}
\tablehead{
\colhead{Parameter} & 
\colhead{Value} &
\colhead{Note/Reference}}
\startdata
SN name   & SN~2008jb & CBET~2771 \\
RA~(J2000)  & 03$^h$51$^m$44$\fs$66 &   \\
DEC~(J2000) & $-38$$\degr$27$\arcmin$00$\farcs$1 &  \\
Spectroscopic Type & II & broad H$\alpha$ in spectrum \\
$E(B-V)_{\rm MW} $ & 0.009~mag & Schlegel et al. (1998) \\
$E(B-V)_{\rm host} $ & $0.06\pm 0.02$~mag & from Balmer decrement \\
$\rm HJD_{exp}$ & 2454782.0 & explosion time \\
$V_{\rm max}/I_{\rm max}$ & $13.59/13.38$~mag & $V$ and $I$ at maximum \\
$M_{V,\rm max}$ & $-16.52$~mag & absolute $V$ mag at maximum \\
$(V-I)_{0,\rm max}$ & $0.13$~mag & unreddened color at maximum \\ 
$V_{\rm mid}/I_{\rm mid}$ & $14.23/13.65$~mag & $V$ and $I$ at mid plateau \\
$M_{V,\rm mid}$ & $-15.87$~mag & absolute $V$ mag at mid plateau\\
$(V-I)_{0,\rm mid}$ & $0.50$~mag & unreddened color at mid plateau\\ 
Duration of ``plateau'' & $\sim 95$~days & from $I$ \\ 
Linear decline slope ($ < 100$~days) &  0.013/0.007~mag~day$^{-1}$ &  $V$ and $I$  \\ 
Linear decline slope ($ > 100$~days) &  0.013~mag~day$^{-1}$ &  $V$  \\ 
\enddata
\end{deluxetable}

%%%%%%%%%%%%%%%%%%%%%%%%%%%%%%%%%%%%%%%%%%%%%%%%

\begin{deluxetable}{lcl}
\tablecolumns{3}
\tablewidth{0pt}
\tablecaption{SN~2008jb host galaxy properties}
\tablehead{
\colhead{Parameter} & 
\colhead{Value} &
\colhead{Note/Reference}}
\startdata
Name   & ESO~302$-$14 &  \\
RA~(J2000)  & 03$^h$51$^m$40$\fs$8 & Paturel et al. (2003) \\
DEC~(J2000) & $-38$$\degr$27$\arcmin$12$\farcs$4 &  Paturel et al. (2003)\\
Morphological Type & IB(s)m & RC3 \\
Heliocentric velocity & $872$~km s$^{-1}$ & Koribalski et al. (2004)\\
Distance modulus & 29.91~mag & using $d_{\rm flow}=9.6$~Mpc \\
$B_{\rm total}$    & $14.84 \pm 0.09$~mag & Lauberts \& Valentijn (1989) \\
$R_{\rm total}$    & $14.54 \pm 0.09$~mag & Lauberts \& Valentijn (1989) \\
$M_{B}$ & $-15.33$~mag & absolute $B$ mag \\
$(B-R)_{0}$ & $0.20$~mag & unreddened color \\
Oxygen abundance~1 & $8.21\pm 0.03$ & PP04~O3N2 method \\
Oxygen abundance~2 & $8.13\pm 0.03$ & PP04~N2 method  \\
Star formation rate & 0.03 $\rm M_{\odot}$~yr$^{-1}$& Lee et al. (2009) \\
Stellar mass & $4.1\times 10^7$~M$_{\odot}$ & M/L from Bell \& de~Jong (2001) \\
\ion{H}{1} mass & $2.5\times 10^8$~M$_{\odot}$ & Meurer et al. (2006) 
\enddata
\end{deluxetable}

%%%%%%%%%%%%%%%%%%%%%%%%%%%%%%%%%%%%%

\clearpage

\begin{deluxetable}{ccccccc}
\tablewidth{0pt}
\tablecaption{Results from black-body fits to optical and mid-IR data}
\tablehead{
\colhead{Epoch$^{\rm a}$} & 
\colhead{L$_{\rm hot}$} & 
\colhead{R$_{\rm hot}$} & 
\colhead{T$_{\rm hot}$} &
\colhead{L$_{\rm warm}$} & 
\colhead{R$_{\rm warm}$} & 
\colhead{T$_{\rm warm}$} \\
\colhead{(days)} & 
\colhead{($10^6$~L$_{\sun}$)} & 
\colhead{(AU)} & 
\colhead{(K)} &
\colhead{($10^6$~L$_{\sun}$)} & 
\colhead{(AU)} & 
\colhead{(K)}
}
\startdata
286.4 & 4.57 & 9.2 & 6000 & 2.77 & 765 & 581 \\
324.0 & 3.01 & 7.5 & 6000 & 1.07 & 633 & 579 \\
433.9 & 0.58 & 3.3 & 6000 & 0.50 & 237 & 680 \\
\enddata
\tablenotetext{a}{HJD - 2454789.7}
\tablecomments{T$_{\rm hot}$ is fixed in all epochs at 6000~K.}
\end{deluxetable}

%%%%%%%%%%%%%%%%%%%%%%%%%%%%%%%%%%%%%%%%%

\end{document}